\documentclass[10pt,a4paper]{article}
\PassOptionsToPackage{table}{xcolor}
\usepackage{Risex-Lab-Paper}
\labsetup{
  name={RISE-X Lab},
  shortname={RISE-X Lab},
  tagline={School of Computer Science,\\Shanghai Jiao Tong University},
  report={},
  date={October 9, 2026},
  website={},
  paper-type={},
  accent={294D73},
  panel={F1F3F6},
  masthead={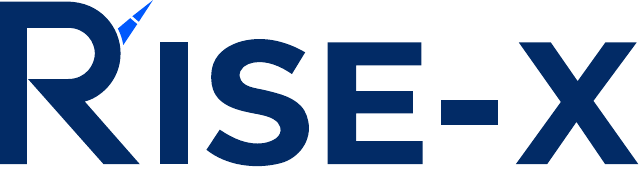},
  masthead-width={25mm}
}

\labsetup{
  text-width={6.2in},
  title-size={17.5},
  title-leading={27},
  running-title-size={11},
  running-title-leading={0},
  date-size={10},
  date-leading={12}
}

\usepackage{amsmath,amsfonts,bm}

\def\eqref#1{equation~\ref{#1}}

\def\1{\bm{1}}

\DeclareMathAlphabet{\mathsfit}{\encodingdefault}{\sfdefault}{m}{sl}
\SetMathAlphabet{\mathsfit}{bold}{\encodingdefault}{\sfdefault}{bx}{n}

\usepackage{algorithm}
\usepackage{algpseudocode}
\usepackage{multirow}
\usepackage{pifont}
\usepackage{pgfplots}
\pgfplotsset{compat=1.18}
\usepgfplotslibrary{groupplots}
\usepackage{listings}
\usepackage{placeins}
\DeclareRobustCommand{\hbfull}{\tikz[baseline=-0.55ex]\fill (0,0) circle (0.75ex);}
\DeclareRobustCommand{\hbhalf}{\tikz[baseline=-0.55ex]{\draw[line width=0.5pt] (0,0) circle (0.75ex); \fill (0,0.75ex) arc (90:270:0.75ex) -- cycle;}}
\DeclareRobustCommand{\hbnone}{\tikz[baseline=-0.55ex]\draw[line width=0.5pt] (0,0) circle (0.75ex);}

\newcommand{\code}[1]{\texttt{#1}}

\title{\texorpdfstring{%
    \raisebox{-0.10em}
{\includegraphics[height=1em]{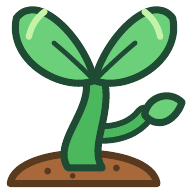}}%
    \hspace{0.20em}%
  }{}%
  SPROUT: Scaling Agentic Reinforcement Learning for Long-Horizon Software Engineering}
\author{\bfseries
Mengyang Liu\textsuperscript{1,2} \quad
	Weiming Wang\textsuperscript{1} \quad
	Haixin Nan\textsuperscript{3} \quad
	Yu Wu\textsuperscript{1} \quad Peize Li\textsuperscript{4} \quad Shihong Huang\textsuperscript{1}
	\\
	\textbf{
	Xuning He\textsuperscript{1} \quad
	Xue Jiang\textsuperscript{5} \quad
	Yihong Dong\textsuperscript{1,\Letter}
	}
}
\affiliations{
\textsuperscript{1}Shanghai Jiao Tong University \quad
\textsuperscript{2}Amazon \quad
\textsuperscript{3}Nanjing University 
\\
\textsuperscript{4}King's College London \quad
\textsuperscript{5}Peking University
}
\paperlinks{\href{https://github.com/RISE-X-Lab/SPROUT}{Code}\enspace /\enspace \href{https://rise-x-lab.github.io/SPROUT/}{Project Page}}
\correspondence{Correspondence: \href{mailto:dongyh@sjtu.edu.cn}{dongyh@sjtu.edu.cn}}
\hypersetup{pdftitle={Scaling Agentic Reinforcement Learning for Long-Horizon Software Engineering},pdfauthor={Mengyang Liu et al.}}

\begin{document}
\makelabtitle{%
Reinforcement learning (RL) has driven substantial progress in coding agents by relying on verified successful trajectories as learning signals. However, such trajectories are scarce in long-horizon software engineering tasks, where a minor error can invalidate the entire rollout.  Repeatedly restarting from the initial state is commonly used yet inefficient, and resetting only the context with working directory misses state changes such as installed dependencies. Therefore, branching from intermediate steps to preserve earlier progress requires restoring both the agent context and execution environments. In this paper, we present \textbf{SPROUT} (\textbf{S}na\textbf{P}shot-based \textbf{R}oll\textbf{OUT}), an agentic RL rollout framework that scales long-horizon coding tasks by resuming and branching from intermediate states. Specifically, SPROUT was built on top of a lightweight infrastructure that checkpoints both the filesystem and the agent context after each assistant turn. A verifier-guided backtracking policy then selects where to resume a failed attempt and how to proceed. Extensive experiments on long-horizon coding tasks show that SPROUT achieves up to 23.6\% improvement over pass@k across various LLMs, while using $32$--$52\%$ of their tokens. For RL training of Qwen3.8-27B, SPROUT stably outperforms standard GRPO by $13.9$ points on three long-horizon coding benchmarks while cutting the time per training step by $39.5\%$. SPROUT thus makes RL on long-horizon coding tasks both more effective and cheaper to scale.
}
\authornote{*}{Work done during research internships at RISE-X Lab, School of Computer Science, Shanghai Jiao Tong University.}
\authornote{$\dagger$}{Work conducted outside of the authors’ roles at Amazon.}
\authornote{}{\Letter: Corresponding author.}

\section{Introduction}
\label{sec:introduction}
Coding agents are becoming capable of completing increasingly long tasks in
real software environments, such as resolving issues in large repositories
\citep{jimenez2024swebench,yang2024sweagent,datacurve2026deepswe}. For such
tasks, reinforcement learning with verifiable rewards (RLVR) has become the de
facto training paradigm \citep{shao2024grpo,he2026bpo,wu2026spark}: the agent
attempts a task in an executable environment, a verifier such as the task's
test suite checks the final result, and the policy is updated from these
outcomes. Such training is especially expensive on long-horizon tasks. A single attempt
may read a large, unfamiliar codebase, install dependencies, build artifacts,
and run test suites, and it can take hours to finish.

The longer the task, the harder it is to learn from. The verifier reward is
usually binary, so an attempt receives credit only if its final solution
passes. Over a long attempt, one wrong decision anywhere,
such as an incorrect patch after hours of correct setup, is enough to make the
reward zero, and successful attempts become rare as the horizon grows. In
group-relative methods such as GRPO \citep{shao2024grpo}, a task whose attempts
all fail has zero advantage for every attempt and contributes no gradient. The
expensive rollouts spent on the hardest tasks thus produce no learning signal,
even when most of their steps were correct.

Two strategies are commonly used to obtain more positive signal. The first
samples more attempts per task \citep{brown2024monkeys,li2022alphacode}, which
multiplies an already high cost: every new attempt redoes its exploration and
setup from scratch and can repeat the same mistake. The second branches from
intermediate steps of earlier attempts. In a coding environment, this requires
three things. The restored environment must match the agent's context at the
branch point, including the files the agent edited, the packages it installed,
and the artifacts it built anywhere in the system. Checkpoints must be cheap
enough to take at every step, since any step may turn out to be the right place
to branch. The branch point itself must also be chosen well. Existing methods
miss at least one of these requirements. Tree search and reflection methods
\citep{yao2023tot,zhou2023lats,shinn2023reflexion}, as well as CBPO
\citep{wang2026cbpo}, branch only within the conversation. Among RL methods
that restore sandbox states, Shepherd \citep{yu2026shepherd} versions only the
agent's workspace, BPO \citep{he2026bpo} snapshots containers only at a few
selected steps, and both rely on layered container snapshots with hard limits
on depth. Full virtual machine snapshots are complete but must store guest
memory at every step, making the storage requirement unacceptable on long horizon tasks. To choose branch points, BPO uses policy entropy without
consulting the verifier, and Shepherd delegates the choice to a stronger
external model. Their RL experiments focused on tasks normally take 8 - 25 steps, leaving long horizon unexplored.

\begin{figure}[t]
\centering
\includegraphics[width=\linewidth]{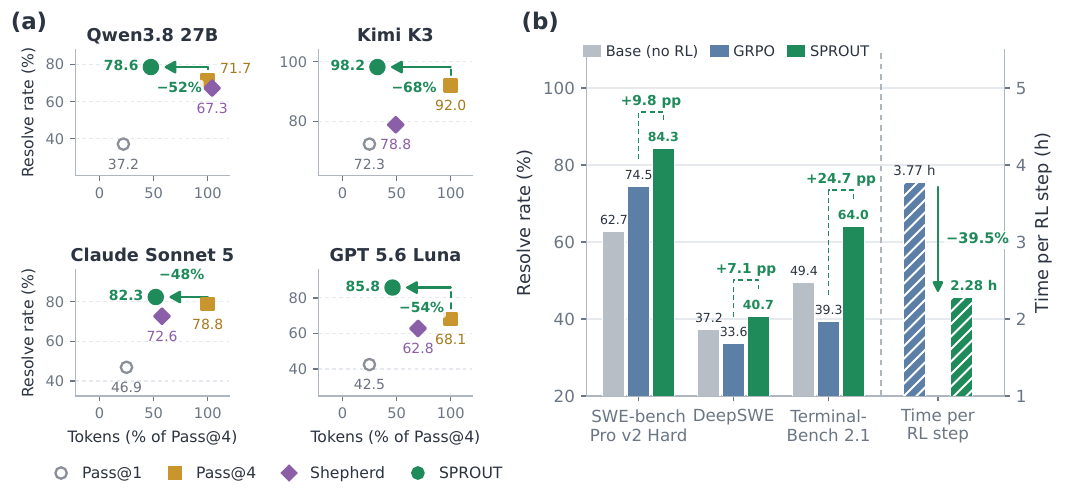}
\caption{\textbf{SPROUT: branch from checkpoints instead of restarting.} Resuming a failed rollout from the step where it went wrong beats resampling it from scratch, at inference time and in RL.
\textbf{(a)}~DeepSWE, four models: SPROUT beats Pass@4 on every model while spending 48--68\% fewer tokens; Shepherd and entropy-based branching do not.
\textbf{(b)}~RL with  Qwen3.8-27B: training on SPROUT rollouts improves over GRPO by 7--25~pp on three mainstream agentic benchmarks. Each RL step is 40\% faster and cheaper than sampling from the beginning.}
\label{fig:sprout-intro}
\end{figure}

\begin{figure}[t]
\centering
\includegraphics[width=\linewidth]{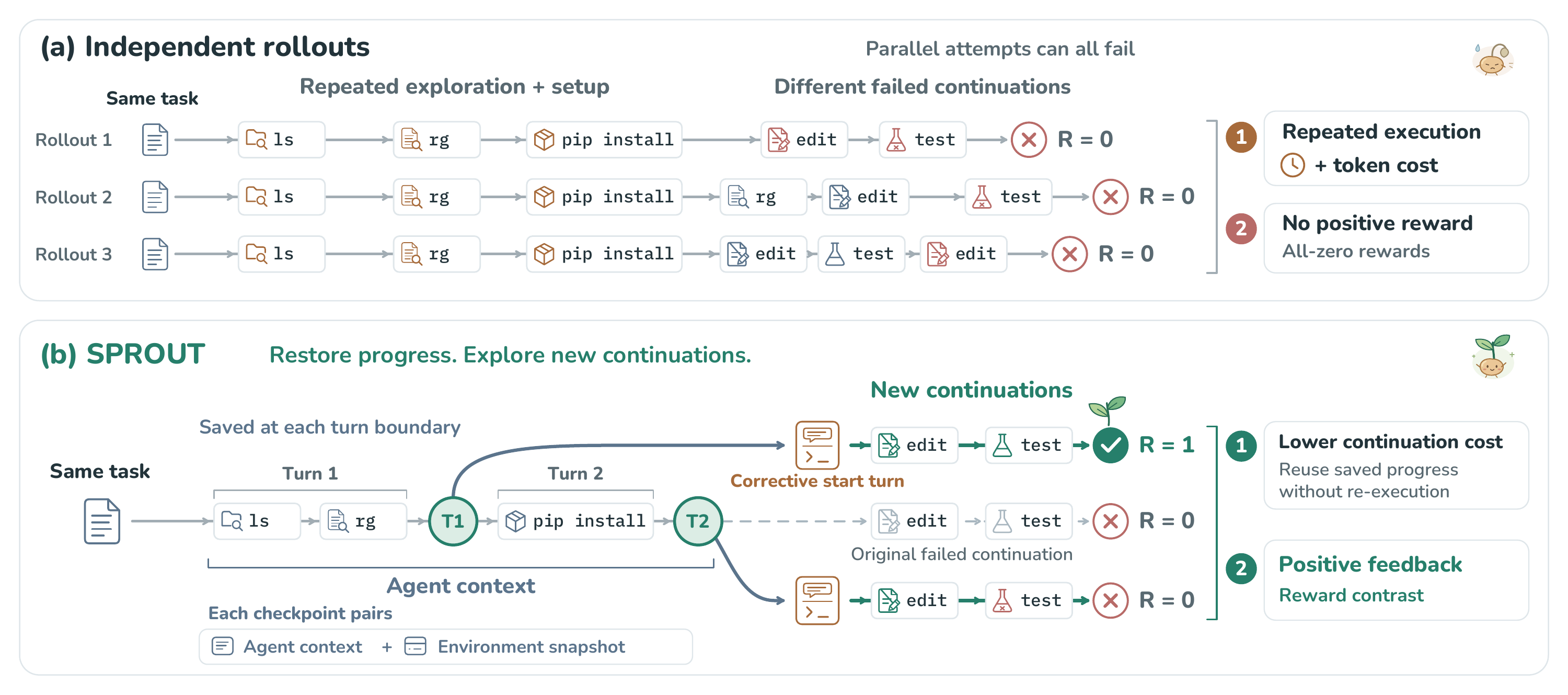}\\[4pt]
\caption{\textbf{SPROUT reuses failed rollouts for efficient exploration.}
(a) Independent rollouts repeat costly prefixes.
(b) Guided by verifier feedback, the same agent policy resumes from paired context and VM checkpoints to explore new continuations.}
\label{fig:sprout-reuse}
\end{figure}

\begin{figure}[t]
\centering
\includegraphics[width=\linewidth]{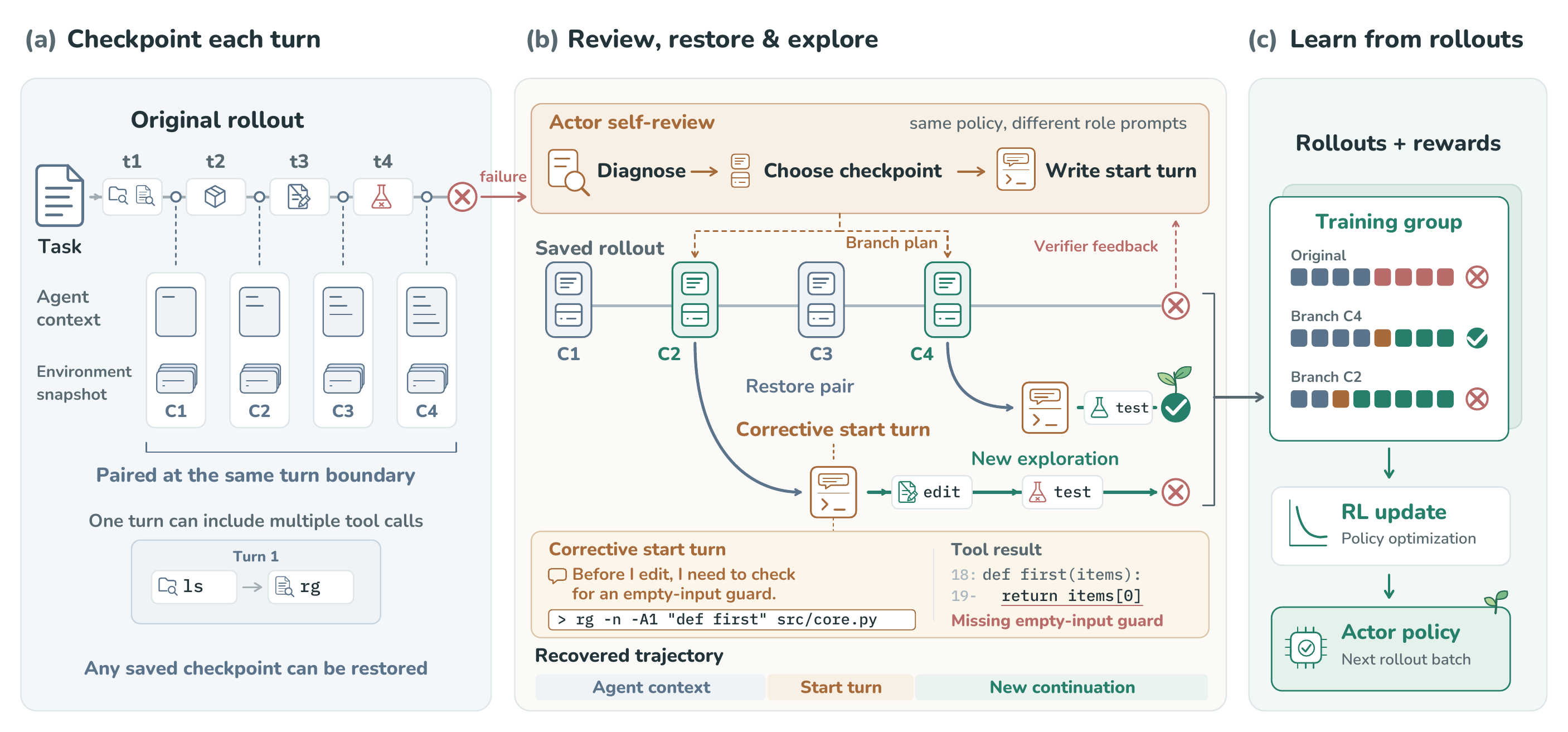}
\caption{\textbf{SPROUT workflow: checkpoint, branch, and learn.}
(a)~After each turn, the execution layer saves a filesystem checkpoint
paired with the matching agent context, so any eligible boundary can be
restored in a fresh sandbox.
(b)~When all root attempts fail, the actor, prompted as analyst and reviewer,
reads the failures together with verifier feedback, selects boundaries, and
writes a starting turn for each branch. Each branch restores its paired
state, executes the starting turn, and continues under the actor policy.
(c)~Roots and branches, with their verified rewards, form training groups;
the review context is removed from the learner's input.}
\label{fig:sprout-workflow}
\end{figure}

In this paper, we present \textbf{\textsc{SPROUT}} (\textbf{S}na\textbf{P}shot-based \textbf{R}oll\textbf{OUT}),
an agentic RL rollout framework that scales long-horizon coding tasks by resuming and branching from intermediate states. Sprout is composed of three components:
1) a restorable execution layer runs each attempt in a microVM
\citep{agache2020firecracker} and take a snapshot of its filesystem incrementally after
every assistant turn, pairing each checkpoint with the matching agent context.
2) a verifier-guided backtracking
policy decides where to branch (Figure~\ref{fig:sprout-workflow}): the actor itself reviews failed trajectories
with the verifier's feedback, chooses where to resume, and proposes the first
action of the new continuation, so the recovered trajectories do not import
knowledge from a stronger model. 3) a GRPO-based training procedure learns from
the branched rollouts and distills the guidance used during search into the
base policy.

We evaluate \textsc{SPROUT} on various mainstream long-horizon agentic coding benchmarks, i.e., DeepSWE, Terminal-Bench 2.1, and SWE-bench Pro
v2 Hard. Compared with a single attempt per task, \textsc{SPROUT} raises the
resolve rate on DeepSWE from $72.3\%$ to $98.2\%$ with Kimi K3, from $46.9\%$ to
$82.3\%$ with Claude Sonnet 5, and from $42.5\%$ to $85.8\%$ with GPT 5.6 Luna,
exceeding four independent attempts while using $32$--$52\%$ of
their tokens. For RL, we train Qwen3.8-27B with
\textsc{SPROUT} on a subset of SWE-bench Rebench v2. Relative to the base model, training raises the resolve
rate from $62.7\%$ to $84.3\%$ on
SWE-bench Pro v2 Hard, from $37.2\%$ to $40.7\%$ on DeepSWE, and from $49.4\%$ to $64.0\%$ on Terminal-Bench 2.1,
ahead of standard GRPO on the same tasks ($74.5\%$, $33.6\%$, and $39.3\%$)
while cutting the 39.5\% of the time per training step.

\section{Related Work}

\paragraph{Long-Horizon Coding Agents.}
Coding agents solve software engineering tasks by interleaving reasoning with tool calls in interactive environments~\citep{yao2022react,dong2024selfcollaboration,wang2024executable}.
Frameworks such as Mini-SWE-agent~\citep{yang2024swe}, OpenHands~\citep{wang2025openhands}, OpenCollab~\citep{hsu2026opencollab}, and Claude Code~\citep{anthropic2025claudecode} let them navigate repositories, edit code, and validate changes through execution, and benchmarks such as SWE-bench Verified~\citep{openai2024swebenchverified}, SWE-bench Pro~\citep{deng2025swe}, DeepSWE~\citep{huang2026deepswe}, and Terminal-Bench 2.1~\citep{merrill2026terminal} evaluate them on increasingly long-horizon tasks.
A single attempt can span tens to hundreds of tool calls and fail on a single error, after which it is usually discarded and resampled from scratch, redoing progress such as understanding the repository and localizing the root cause.
\textsc{SPROUT} instead resumes failed attempts from intermediate states and reuses their prior progress.

\paragraph{Tree Search and Backtracking.}
Repeated sampling increases task coverage but restarts every sample from the initial state~\citep{brown2024large, ROCODE}.
Tree-search methods instead expand and backtrack over intermediate states~\citep{yao2023tree,hao2023reasoning,zhou2023language}, much as Go-Explore first returns to a visited state and then explores~\citep{ecoffet2021first}, while reflection methods correct errors in later attempts using verbal or tool feedback~\citep{shinn2023reflexion,gou2024critic}.
In software engineering, SWE-Search~\citep{antoniades2025swe} applies Monte Carlo tree search to issue resolution, and Shepherd~\citep{yu2026shepherd} delegates the choice of where to intervene over reversible traces to a stronger external model.
Because agents edit files, install packages, and build artifacts, branching only within the conversation, as tree search and reflection methods do, leaves the environment inconsistent with the agent's context, and Shepherd versions only the agent's workspace.
\textsc{SPROUT} restores the full filesystem together with the matching agent context and lets the actor itself, rather than a stronger model, use verifier feedback to choose where to branch and how to continue.
\begin{table}[t]
\centering
\footnotesize
\renewcommand{\arraystretch}{1.25}
\caption{\textbf{Comparison with representative approaches.} \hbfull\ full,
\hbhalf\ partial, \hbnone\ no support; -- not applicable. \emph{Consistent}:
the agent context and the environment are restored to the same step.
\emph{Full scope}: the restored environment covers the whole disk, not only a
workspace. \emph{Per-step}: every tool call of a long trajectory yields a
restorable checkpoint. \emph{Compact}: checkpoints keep incremental disk
changes, snapshot is cheap enough for taking snapshot for each tool call. \emph{Failure-aware}: branching uses verifier
feedback on the failed attempt. \emph{Self-guided}: the trained policy, not a
stronger model, makes the branching decisions.}
\label{tab:method-comparison}
\begin{tabularx}{\linewidth}{l*{6}{>{\centering\arraybackslash}X}}
\toprule
& \multicolumn{4}{c}{\textbf{State restoration}} & \multicolumn{2}{c}{\textbf{Branch selection}} \\
\cmidrule(lr){2-5}\cmidrule(lr){6-7}
\textbf{Method} & \textbf{Consistent} & \textbf{Full scope} & \textbf{Per-step} & \textbf{Compact} & \textbf{Failure-aware} & \textbf{Self-guided} \\
\midrule
\multicolumn{7}{@{}l}{\emph{Retry with feedback}} \\
\quad Reflexion & \hbnone & -- & -- & -- & \hbfull & \hbfull \\
\addlinespace[2pt]
\multicolumn{7}{@{}l}{\emph{Branching for RL}} \\
\quad TreeRL / CBPO & \hbhalf & \hbnone & \hbnone & -- & \hbnone & \hbfull \\
\quad BPO & \hbfull & \hbfull & \hbhalf & \hbfull & \hbnone & \hbfull \\
\quad Shepherd & \hbfull & \hbhalf & \hbhalf & \hbfull & \hbhalf & \hbnone \\
\addlinespace[2pt]
\multicolumn{7}{@{}l}{\emph{Checkpointing substrates}} \\
\quad Docker & \hbhalf & \hbfull & \hbhalf & \hbfull & -- & -- \\
\quad MicroVM snapshots & \hbhalf & \hbfull & \hbfull & \hbnone & -- & -- \\
\midrule
\rowcolor{gray!12}
\textbf{\textsc{SPROUT}} (ours) & \hbfull & \hbfull & \hbfull & \hbfull & \hbfull & \hbfull \\
\bottomrule
\end{tabularx}
\end{table}
\paragraph{Stateful Sandboxing and Agentic RL.}
In sparse-reward agent tasks, groups whose attempts all fail are common and yield no gradient under group-relative methods such as GRPO~\citep{shao2024deepseekmath}\nocite{RL-PLUS}.
Several RL methods therefore branch from intermediate states, including VinePPO~\citep{kazemnejad2024vineppo} for advantage estimation, TreeRL~\citep{hou2025treerl} for process supervision, SPARK~\citep{wu2026spark} at critical states identified by the agent's own decision signals, CBPO~\citep{wang2026contrastive} for contrastive branching, and BPO~\citep{he2026branching} with container snapshots.
However, most of these methods branch only within the conversation, which in coding environments leaves the environment inconsistent with the agent's context, while BPO can resume only from a few steps selected by policy entropy without consulting the verifier, so neither approach supports consistent branching from an arbitrary tool boundary.
Among checkpointing mechanisms~\citep{wang2026fork,li2025agentgit,xu2025toward}, layered container snapshots~\citep{merkel2014docker} have hard limits on depth and full microVM snapshots~\citep{agache2020firecracker} must store guest memory at every step.
\textsc{SPROUT} checkpoints the full filesystem incrementally after every tool call and pairs each checkpoint with the matching agent context, so any tool boundary can be restored with low storage overhead and without re-executing the prefix, enabling more efficient exploration to obtain learning signals.

\section{\textsc{SPROUT}}
\label{sec:pair}

\textsc{SPROUT} is a rollout framework for long-horizon agentic tasks, it allows resuming execution
from any recorded intermediate state of an earlier attempt and explore
alternative continuations from that state. We first formulate the rollout
collection problem and give an overview of the framework. We then describe
the execution layer that enables resumption and the verifier-guided
backtracking policy that determines where to resume and how to continue.
Finally, we present policy optimization with the resulting rollouts.

\subsection{Problem Formulation}
\label{sec:problem}

A software engineering task $x=(q_x,e_0)$ consists of a description $q_x$ and
an initial environment $e_0$. The agent interacts with the environment in
\emph{turns}. In turn $t$, the actor policy $\pi_\theta$ produces an assistant
message $a_t$ consisting of reasoning followed by a sequence of tool calls
$(a_t^{1},\ldots,a_t^{n_t})$. The harness executes these calls in order and
returns their results $o_{t+1}=(o_{t+1}^{1},\ldots,o_{t+1}^{n_t})$ as the
observation of the turn. After $t$ completed turns, the agent is at
\emph{boundary} $t$ with joint state $s_t=(e_t,h_t)$, where $e_t$ is the
environment after all tool calls of the preceding turn have completed and
$h_t$ is the context retained by the agent harness. Each turn follows
\begin{equation}
 a_t\sim\pi_\theta(\cdot\mid h_t),\qquad
 (e_{t+1},o_{t+1})\sim P_x(\cdot\mid e_t,a_t),\qquad
 h_{t+1}=\operatorname{Update}(h_t,a_t,o_{t+1}),
 \label{eq:interaction}
\end{equation}
where $P_x$ is the transition and observation distribution induced by
executing all tool calls of $a_t$, and $\operatorname{Update}$ is the
harness's context-management procedure. Boundaries lie between turns rather
than between individual tool calls. Within a turn, the actor has already
committed to all of its calls, so an environment state reached partway
through the turn has no matching agent context from which the policy could
continue. A completed attempt with $T$ turns is a trajectory
$\tau=(h_0,a_0,o_1,\ldots,a_{T-1},o_T)$, and $\tau_{\le t}$ denotes its
prefix up to and including $o_t$. The task verifier returns
\begin{equation}
 (R_x(\tau),d_x(\tau))=\mathcal{V}_x(\tau),\qquad R_x(\tau)\in\{0,1\},
 \label{eq:verification}
\end{equation}
where $R_x$ is the binary terminal reward and $d_x(\tau)$ is diagnostic
feedback such as failing tests. We call a trajectory with $R_x(\tau)=1$ a
\emph{positive trajectory}.

\subsection{Framework Overview}
\label{sec:overview}

The central capability of \textsc{SPROUT} is \emph{local access}: a search
procedure can return to an earlier joint state and resume execution from it
\citep{weisz2021query,yin2022efficient}. During an attempt $\tau_i$, the
framework records a handle $c_{i,t}$ at each turn boundary $t$, pairing the
environment checkpoint with the matching agent context. A valid handle
instantiates a branch state
\begin{equation}
 e'=\operatorname{Restore}(c_{i,t}),\qquad
 h'=\operatorname{Recover}(c_{i,t}),\qquad s'=(e',h'),
 \label{eq:local-access}
\end{equation}
where $\operatorname{Restore}$ brings a fresh sandbox to the checkpointed
environment and $\operatorname{Recover}$ returns the context at the same
boundary, so $e'$ and $h'$ both correspond to the executed prefix
$\tau_{i,\le t}$. A new continuation $\eta$ sampled from $s'$ yields the
attempt
\begin{equation}
 \tau'=\tau_{i,\le t}\oplus\eta.
 \label{eq:protocols}
\end{equation}
The prefix is inherited rather than re-executed: its $t$ turns remain part
of $\tau'$, and only $\eta$ requires new execution.

This interface separates execution from search. The execution layer makes
recorded states available for isolated continuations, and a replaceable
search policy decides which states to revisit and whether to give guidance when start the continuation; randomized branching, heuristic selection, and tree
search all fit the same interface. An archive $\mathcal{A}_x$ stores the
resulting execution tree for task $x$, including checkpoint handles, parent
links, contexts, and verifier feedback, and $\mathcal{T}_x$ denotes the
multiset of its completed attempts. New outcomes in the archive inform
further search, and completed attempts are organized into training groups.

\subsection{Restorable Execution States}
\label{sec:restorable-states}

\paragraph{Consistent Continuation Boundary.}
\label{sec:context-recovery}
Local access requires the environment and the agent context to describe the same
point of an attempt. Boundary $t$ of $\tau_i$ is \emph{eligible},
$t\in\mathcal{E}(\tau_i)$, if a filesystem checkpoint was taken after all
tool calls of turn $t$ completed and the conversation prefix can be
recovered up to and including the observation $o_t$. Search branches only
from eligible boundaries, so every continuation starts from the observations
and execution state of its selected prefix. Any guidance for a new
continuation is stored separately from that prefix.

\paragraph{Filesystem restoration under the harness contract.}
\label{sec:persistence}
Each attempt runs in its own microVM~\citep{agache2020firecracker}, so a
checkpoint covers the whole guest filesystem, including dependencies and
build artifacts outside the working directory. Our harness executes every
tool call as a synchronous \code{shell} command and disallows background
processes, so by the end of a turn no process started by the agent is still
running, and all state that persists across turns lives on disk.
$\operatorname{Restore}$ therefore boots a fresh microVM from the saved
filesystem and reinitializes the tool runtime, and guest memory never needs
to be captured. Together with the recovered context, this gives the
continuation the full state of its prefix under the stated harness
constraints.

\paragraph{Incremental checkpoints and isolated branches.}
\label{sec:branch-isolation}
After each turn, the execution layer records the changes to the VM disk as
an incremental block-level OverlayBD layer. Consecutive checkpoints share all
unchanged content, so the storage a checkpoint adds grows with the blocks
written during the turn rather than with the size of the environment. Each
restored branch writes to its own copy-on-write layer over the saved history,
so several continuations can inherit the same progress without affecting one
another or their parent. Incremental storage and isolated writes together
make it practical to checkpoint every turn of a long trajectory.

\subsection{Verifier-guided Backtracking Policy}
\label{sec:method}

Restoration makes earlier states available; the search policy decides which
of them are worth continuing from. We instantiate it with verifier-guided
backtracking, in which the actor $\pi_\theta$ also analyzes failures and
plans branches under separate role prompts. The roles share parameters, and
only the analysis and planning roles see feedback from completed attempts,
so no separate analyst or reviewer model is involved.

\paragraph{From failure evidence to branch decisions.}
For each newly failed attempt $\tau_i$, the policy writes an analysis $z_i$
of why the attempt failed, which of its progress can be reused, and which
eligible boundaries could plausibly lead to a successful continuation. These
candidates are hypotheses about where recovery is possible; they need not
identify the first incorrect turn. In round $r$, a reviewer prompt reads the
archive, now including these analyses, and plans up to $K_r$ branches:
\begin{equation}
 \{(i_j,t_j,a'_j)\}_{j=1}^{K}
 \sim\pi_\theta(\cdot\mid\mathrm{review},x,\mathcal{A}_x),
 \qquad K\le K_r,\quad t_j\in\mathcal{E}(\tau_{i_j}).
 \label{eq:selection}
\end{equation}
Each decision specifies a parent attempt $i_j$, a boundary $t_j$, and a
starting turn $a'_j$ that replaces the parent's turn $t_j$. Like any actor
turn, $a'_j$ consists of reasoning in the actor's voice followed by one or
more tool calls. Because the parent is chosen per branch, the reviewer can
return to an older attempt instead of extending a dead end, or build on a
failed branch from an earlier round. The archive also records earlier
starting turns and their outcomes, which helps the reviewer avoid directions
that have already failed.

\paragraph{Executing and evaluating a continuation.}
For each decision, the framework restores $(e'_j,h'_j)$ from $c_{i_j,t_j}$ in
a fresh sandbox and executes the tool calls of $a'_j$ in order, exactly as in
an ordinary turn, obtaining the observation $o'_j$. Let
$s_j^+=(e_j^+,h_j^+)$ denote the joint state at the resulting boundary
$t_j+1$, after the harness applies its context update. The actor then
continues without further guidance:
\begin{equation}
 \eta_j\sim\rho_\theta(\cdot\mid s_j^+),\qquad
 \tau'_j=\tau_{i_j,\le t_j}\oplus(a'_j,o'_j)\oplus\eta_j,
 \label{eq:continuation}
\end{equation}
where $\rho_\theta(\cdot\mid s)$ is the distribution over continuations
obtained by running $\pi_\theta$ from $s$ until termination. The verifier
grades the completed attempt, and the archive records its outcome together
with its parent, boundary, and starting turn. A failed continuation thus
contributes both reusable execution progress and new diagnostic evidence to
the next round.

In our rollout-collection experiments, we apply this procedure to tasks whose
initial attempt fails, with $R=2$ rounds of $K_1=4$ and $K_2=3$ branches.
Search stops as soon as a branch passes or after the last round. Because the
procedure uses diagnostic feedback beyond the binary reward, these
experiments measure exploration in a setting where such feedback is
available, not unassisted agent performance.

\subsection{Policy Optimization with Branched Rollouts}
\label{sec:training}

Training reuses the same restoration and repair mechanisms to construct
groups. During search, the reviewer sees diagnostics that never appear in the
actor's ordinary context. Training aims to transfer the resulting corrective
behavior into the actor while accounting for the prefixes that trajectories
in a group share.

\paragraph{Constructing training groups.}
Each task receives a fixed-size group: two independent roots and $K=3$
retained children per root, for eight evaluated attempts. This specializes
the collection framework to a single layer of children. If at least one root
succeeds (TT or TF), the reviewer selects one usable boundary per root, and
that root's children continue from it without an injected starting turn. If
both roots fail (FF), the reviewer uses verifier feedback to rank boundaries
and attaches a corrective direction to each. A starting turn is then written
from each visible prefix and its direction, filtered by a judge, and executed
in the restored environment before the actor continues. A boundary is usable
only if it is eligible and its prefix leaves room for the continuation within
the context budget. Algorithm~\ref{alg:sprout} (Appendix~\ref{app:algorithm})
specifies the oversampling, independent-rollout fallbacks, and task
replacement that keep every group at eight attempts.

\paragraph{Learning without the review context.}
For a guided branch, the learner sees the retained prefix, the starting turn,
and the executed continuation, but not the diagnostic analysis or the
corrective direction. The policy is therefore trained to produce corrective
turns from the context it has during ordinary execution. Removing the review
context from the learner's input does not remove its influence on how the
sample was generated. We therefore treat the objective below as a clipped
surrogate that distills guided behavior into the actor, not as on-policy
optimization over samples from the unassisted actor.

\paragraph{Credit assignment and shared prefixes.}
In training, $\mathcal{T}_x$ denotes the eight retained attempts of task $x$.
Depending on the outcomes of the two roots, we assign
\begin{equation}
 A(\tau)=
 \begin{cases}
 R_x(\tau), & \tau\text{ is a child in an FF group},\\[2pt]
 \dfrac{R_x(\tau)-\bar R_x}{\sigma_x}, & \text{otherwise},
 \end{cases}
 \label{eq:tree-credit}
\end{equation}
where $\bar R_x$ and $\sigma_x$ are the mean and standard deviation of
rewards over $\mathcal{T}_x$, and the normalized advantage is set to zero
when $\sigma_x=0$. Successful FF children thus receive an advantage of one,
and failed FF children receive zero.

A token mask $M(\tau)$ selects the action tokens each attempt contributes to
the loss. Roots contribute the tokens of all their turns, and TT and TF
children contribute only the turns generated after their branch boundaries.
Within an FF family, the first successful child contributes its full
trajectory, and later successful children omit the prefix they share with
it. A recovered success therefore reinforces the progress it inherited, while
successful siblings do not credit the same prefix repeatedly. Observation
tokens are always excluded.

The policy maximizes
\begin{equation}
 \mathcal{J}(\theta)=\mathbb{E}_x\left[
 \sum_{\tau\in\mathcal{T}_x}
 \frac{1}{|M(\tau)|}\sum_{k\in M(\tau)}
 \min\left(
 w_k(\theta)A(\tau),
 \operatorname{clip}(w_k(\theta),1-\epsilon,1+\epsilon)A(\tau)
 \right)\right],
 \label{eq:tree-loss}
\end{equation}
where $y_k$ is the $k$-th token of $\tau$ in the learner's context,
$w_k(\theta)=\pi_\theta(y_k\mid y_{<k})/
\pi_{\theta_{\mathrm{old}}}(y_k\mid y_{<k})$, attempts with empty masks are
skipped, and no KL penalty is used. For guided branches, $w_k$ is not an
importance weight with respect to the diagnostic-conditioned procedure that
generated the sample. This is the surrogate described above, and
Section~\ref{sec:rl} evaluates its effect on the trained actor.
\section{Experiments}
\label{sec:setup}
\label{sec:results}

We evaluate \textsc{SPROUT} with Kimi K3 \citep{DBLP:journals/corr/abs-2607-24653},
Qwen3.8-27B \citep{qwen2026qwen38},
DeepSeek V4.1 Flash \citep{deepseek2026v41flash},
Claude Sonnet 5 \citep{anthropic2026sonnet5}, and
GPT-5.6 Luna \citep{openai2026gpt56} 
on DeepSWE \citep{huang2026deepswe},
Terminal-Bench 2.1 \citep{merrill2026terminal}, and
SWE-bench Pro v2 Hard \citep{deng2025swe}.
We compare \textsc{SPROUT} with independent sampling (pass@k), entropy-based branch sampling,
and Shepherd \citep{yu2026shepherd}.
We measure task resolve rate, recovery rate, execution steps, and
total tokens for rollout collection, and further evaluate
RL training performance.
Detailed setups and further experiments are provided in Appendices~\ref{app:experiment}-\ref{app:prompts}.

\begin{table}[h!]
\centering
\scriptsize
\setlength{\tabcolsep}{3pt}
\renewcommand{\arraystretch}{1.12}

\caption{\textbf{Main experimental comparison on DeepSWE.}
Resolve rate, recovery rate, total execution steps, and total tokens (M).
A dash (--) indicates that Entropy-based sampling is unavailable. Best values are bolded, excluding Pass@1 (Base).}

\label{tab:sampling-matrix}

\begin{tabularx}{\linewidth}{@{}ll*{5}{>{\centering\arraybackslash}X}@{}}
\toprule
\multirow{2}{*}{\textbf{Model}}
& \multirow{2}{*}{\textbf{Metric}}
& \textbf{Pass@1}
& \multirow{2}{*}{\textbf{Pass@4}}
& \textbf{Entropy-}
& \multirow{2}{*}{\textbf{Shepherd}}
& \multirow{2}{*}{\textbf{SPROUT}} \\
& & \textbf{(Base)} & & \textbf{based} & & \\
\midrule

Qwen3.8 27B
& Resolve rate (\%) $\uparrow$
& $37.2$
& $71.7$
& $71.7$
& $67.3$
& $\mathbf{78.6}$ \\

& Recovery (\%) $\uparrow$
& $0.0$
& $54.9$
& $54.9$
& $47.9$
& $\mathbf{64.8}$ \\

& Steps $\downarrow$
& $16{,}708$
& $64{,}136$
& $45{,}549$
& $55{,}280$
& $\mathbf{23{,}174}$ \\

& Tokens (M) $\downarrow$
& $1921$
& $8673$
& $8589$
& $9026$
& $\mathbf{4123}$ \\

\addlinespace[3pt]

Kimi K3
& Resolve rate (\%) $\uparrow$ & $72.3$ & $92.0$ & -- & $78.8$ & $\mathbf{98.2}$ \\
& Recovery (\%) $\uparrow$ & $0.0$ & $71.2$ & -- & $23.2$ & $\mathbf{93.6}$ \\
& Steps $\downarrow$ & $13{,}001$ & $52{,}003$ & -- & $20{,}668$ & $\mathbf{16{,}065}$ \\
& Tokens (M) $\downarrow$ & $1154$ & $4614$ & -- & $2272$ & $\mathbf{1492}$ \\

\addlinespace[3pt]

DeepSeek V4.1 Flash
& Resolve rate (\%) $\uparrow$ & $57.5$ & $82.3$ & -- & $86.7$ & $\mathbf{90.3}$ \\
& Recovery (\%) $\uparrow$ & $0.0$ & $58.3$ & -- & $68.8$ & $\mathbf{77.1}$ \\
& Steps $\downarrow$ & $14{,}533$ & $58{,}372$ & -- & $30{,}143$ & $\mathbf{20{,}230}$ \\
& Tokens (M) $\downarrow$ & $1459$ & $6109$ & -- & $3624$ & $\mathbf{2538}$ \\

\addlinespace[3pt]

Claude Sonnet 5
& Resolve rate (\%) $\uparrow$ & $46.9$ & $78.8$ & -- & $72.6$ & $\mathbf{82.3}$ \\
& Recovery (\%) $\uparrow$ & $0.0$ & $60.0$ & -- & $48.3$ & $\mathbf{66.7}$ \\
& Steps $\downarrow$ & $15{,}362$ & $61{,}450$ & -- & $33{,}395$ & $\mathbf{28{,}592}$ \\
& Tokens (M) $\downarrow$ & $1872$ & $7487$ & -- & $4328$ & $\mathbf{3897}$ \\

\addlinespace[3pt]

GPT 5.6 Luna
& Resolve rate (\%) $\uparrow$ & $42.5$ & $68.1$ & -- & $62.8$ & $\mathbf{85.8}$ \\
& Recovery (\%) $\uparrow$ & $0.0$ & $44.6$ & -- & $35.4$ & $\mathbf{75.4}$ \\
& Steps $\downarrow$ & $6{,}020$ & $24{,}081$ & -- & $13{,}886$ & $\mathbf{8{,}634}$ \\
& Tokens (M) $\downarrow$ & $352$ & $1409$ & -- & $985$ & $\mathbf{652}$ \\

\bottomrule
\end{tabularx}
\end{table}

\begin{table}[t]
\centering
\scriptsize
\setlength{\tabcolsep}{3pt}
\renewcommand{\arraystretch}{1.12}

\caption{\textbf{Experimental comparison across benchmarks for Qwen3.8 27B.}
Resolve rate, recovery rate, total execution steps, and total tokens (M).
Best values are bolded, excluding Pass@1.}
\label{tab:sampling-matrix-qwen38-27b}

\begin{tabularx}{\linewidth}{@{}ll*{5}{>{\centering\arraybackslash}X}@{}}
\toprule
\multirow{2}{*}{\textbf{Benchmark}}
& \multirow{2}{*}{\textbf{Metric}}
& \textbf{Pass@1}
& \multirow{2}{*}{\textbf{Pass@4}}
& \textbf{Entropy-}
& \multirow{2}{*}{\textbf{Shepherd}}
& \multirow{2}{*}{\textbf{SPROUT}} \\
& & \textbf{(Base)}  & & \textbf{based} & & \\
\midrule

\multirow{4}{*}{\textbf{DeepSWE}}

& Resolve rate (\%) & $37.2$ & $71.7$ & $71.7$ & $67.3$ & $\mathbf{78.6}$ \\
& Recovery (\%) & $0.0$ & $54.9$ & $54.9$ & $47.9$ & $\mathbf{64.8}$ \\
& Steps & $16708$ & $64136$ & $45549$ & $55280$ & $\mathbf{23174}$ \\
& Tokens (M) & $1921$ & $8673$ & $8589$ & $9026$ & $\mathbf{4123}$ \\

\addlinespace[3pt]
\midrule
\addlinespace[3pt]

\multirow{4}{*}{\shortstack[l]{\textbf{Terminal-}\\\textbf{Bench 2.1}}}
& Resolve rate (\%) & $49.4$ & $66.3$ & $76.4$ & $79.8$ & $\mathbf{89.9}$ \\
& Recovery (\%) & $0.0$ & $33.3$ & $53.3$ & $60.0$ & $\mathbf{80.0}$ \\
& Steps & $1943$ & $10890$ & $10374$ & $10638$ & $\mathbf{6685}$ \\
& Tokens (M) & $27.65$ & $319.9$ & $112.8$ & $\mathbf{103.9}$ & $123.7$ \\

\addlinespace[3pt]
\midrule
\addlinespace[3pt]

\multirow{4}{*}{\shortstack[l]{\textbf{SWE-bench Pro}\\\textbf{v2 Hard}}}

& Resolve rate (\%) & $62.7$ & $88.2$ & $90.2$ & $78.4$ & $\mathbf{94.1}$ \\
& Recovery (\%) & $0.0$ & $68.4$ & $73.7$ & $42.1$ & $\mathbf{84.2}$ \\
& Steps & $6939$ & $25184$ & $11953$ & $11841$ & $\mathbf{9139}$ \\
& Tokens (M) & $550.8$ & $1918.5$ & $1238.9$ & $1004.0$ & $\mathbf{771.3}$ \\

\bottomrule
\end{tabularx}
\end{table}

\subsection{Sampling Results}
\label{sec:sampling-results}

Table~\ref{tab:sampling-matrix} compares task resolution and resource consumption across sampling methods. Overall,
SPROUT achieves high task coverage while substantially reducing execution and inference costs.
On DeepSWE, SPROUT matches or exceeds Pass@4 across all four models, using 53.5--69.1\% fewer execution
steps and 47.9--67.7\% fewer tokens. The improvement is particularly clear with GPT 5.6 Luna: the
resolve rate increases from 68.1\% to 85.8\%, while token usage decreases from 1,409M to 652M.
With Kimi K3, SPROUT reaches a 98.2\% resolve rate using only 32.3\% of the tokens required by
Pass@4. It also improves over Shepherd by 9.7--23.0 percentage points across the four models while
consistently using fewer steps and tokens. Recovery rates of 64.8--93.6\% further show that failure-
guided backtracking can resolve a substantial fraction of initially unsuccessful tasks. SPROUT also
compares favorably with alternative adaptive sampling strategies. With Qwen3.8-27B, it achieves a
78.6\% resolve rate and a 64.8\% recovery rate, compared with 71.7\% and 54.9\% for entropy-based
sampling, while reducing token usage from 8,589M to 4,123M. Its largest improvement over Pass@4 is obtained with Qwen3.8-27B
on Terminal-Bench 2.1 (Table~\ref{tab:sampling-matrix-qwen38-27b}), where the resolve rate increases
from 66.3\% to 89.9\%. This
gain of 23.6 percentage points is achieved with 38.6\% fewer execution steps and 61.3\% fewer tokens,
together with an 80.0\% recovery rate. As a result, SPROUT maintains competitive task coverage at
much lower cost, demonstrating that SPROUT effectively uses failure feedback to identify recovery
points and achieve a strong balance between task resolution and sampling cost.

\begin{table}[h]
\centering
\scriptsize
\caption{\textbf{Training comparison.} Benchmark columns report success rates
in percent. Red annotations show \textsc{SPROUT} minus Basic GRPO in percentage
points (pp) for success rates, and the relative reduction for time per step.}
\label{tab:training-baselines}
\begin{tabularx}{\linewidth}{@{}l*{4}{>{\centering\arraybackslash}X}@{}}
\toprule
\textbf{Method} & \shortstack{\textbf{SWE-bench Pro v2 Hard}} & \textbf{DeepSWE}
& \textbf{Terminal-Bench 2.1}
& \textbf{Time per step ($\downarrow$)} \\
\midrule
Base (no RL)  & $62.7$ & $37.2$ & $49.4$  & n/a \\
Basic GRPO  & $74.5$ & $33.6$ & $39.3$ & $3.77\,\mathrm{h}$ \\
\textsc{SPROUT} & $\mathbf{84.3}\,\textcolor{red!70!black}{(\uparrow\,9.8\,\mathrm{pp})}$
& $\mathbf{40.7}\,\textcolor{red!70!black}{(\uparrow\,7.1\,\mathrm{pp})}$
& $\mathbf{64.0}\,\textcolor{red!70!black}{(\uparrow\,24.7\,\mathrm{pp})}$
& $\mathbf{2.28}\,\mathrm{h}\,\textcolor{red!70!black}{(\downarrow\,39.5\%)}$ \\
\bottomrule
\end{tabularx}
\par\smallskip
\end{table}

\subsection{RL training}
\label{sec:rl}

We train Qwen3.8-27B on SWE-bench Rebench v2 Filtered,
fine-tuning all parameters, and compare against standard GRPO on the same tasks
(Table~\ref{tab:training-baselines}). Training with \textsc{SPROUT} improves the base model on all three benchmarks:
from $62.7\%$ to $84.3\%$ on SWE-bench
Pro v2 Hard, from $37.2\%$ to $40.7\%$ on DeepSWE, and from $49.4\%$ to
$64.0\%$ on Terminal-Bench 2.1. These gains are $21.6$, $3.5$, and $14.6$
percentage points, respectively. Compared with Basic GRPO, \textsc{SPROUT}'s
success rates are $9.8$, $7.1$, and $24.7$ percentage points higher on the
same three benchmarks, stably outperforming with an average of $13.9$ points.
\textsc{SPROUT} also reduces time per training step from $3.77$ to $2.28$
hours, a $39.5\%$ reduction.

\begin{table}[ht]
\centering
\renewcommand{\arraystretch}{1.2}
\caption{\textbf{Ablations of \textsc{SPROUT} on DeepSWE using GPT-5.6 Luna}.
Each variant changes one component and keeps the rest of the collection
procedure fixed. Res.\ is the resolve rate (\%) and $\Delta$ the change relative
to \textsc{SPROUT}. Branches, Steps (tool calls), and Tok.\ (M) count the branching stage only, excluding the root
run.}
\label{tab:ablation}
\footnotesize
\begin{tabularx}{\linewidth}{l>{\raggedright\arraybackslash}X*{5}{c}}
\toprule
\textbf{Variant} & \textbf{Change} & \textbf{Res.\,$\uparrow$} & $\boldsymbol{\Delta}$ & \textbf{Branches\,$\downarrow$} & \textbf{Steps\,$\downarrow$} & \textbf{Tok.\,$\downarrow$} \\
\midrule
\rowcolor{gray!12}
\textbf{\textsc{SPROUT}} (tool-call hint) & full method & $\mathbf{85.8}$ & -- & $325$ & $2{,}839$ & $261.9$ \\
\multicolumn{7}{l}{\emph{Continuation guidance}} \\
\quad Rollback, no hint & no injected guidance & $49.6$ & $-36.2$ & $\mathbf{183}$ & $\mathbf{1{,}755}$ & $\mathbf{209.8}$ \\
\multicolumn{7}{l}{\emph{Branch selection}} \\
\quad Reward-only review & binary outcome only & $53.1$ & $-32.7$ & $153$ & $2{,}296$ & $274.2$ \\
\bottomrule
\end{tabularx}
\end{table}

\subsection{Ablations}
\label{sec:ablations}

Table~\ref{tab:ablation} isolates the components of \textsc{SPROUT}; each
variant changes one component and keeps the rest of the collection procedure
fixed.
\emph{Continuation guidance.} \emph{Rollback, no hint} resumes from the
selected checkpoint without any injected guidance, which measures what the
reviewer's starting turn adds beyond restoration alone.
\emph{Branch selection.} \emph{Reward-only review} hides the verifier's
diagnostics and keeps only the binary outcome, which measures how much the
diagnostic feedback contributes to choosing where and how to branch.
Both components matter. Without the injected hint, the resolve rate falls from
$85.8\%$ to $49.6\%$. Hiding the verifier's diagnostics lowers it to $53.1\%$. The reviewer's
guidance and the verifier's diagnostics therefore both raise the success of
branching and make the search cheaper.

\begin{figure}[h]
  \centering
  \includegraphics[width=0.8\linewidth]{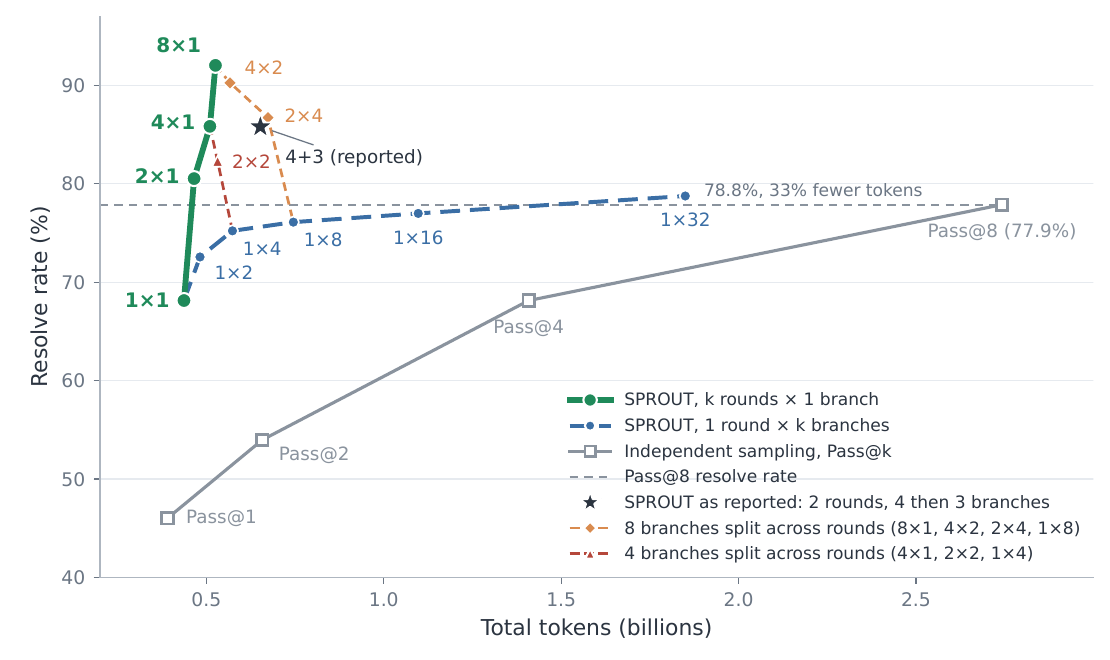}
  \caption{
    \textbf{Branch scaling and round allocation.}
    Resolve rate on DeepSWE using GPT-5.6 Luna, where
    labels denote review rounds $\times$ branches per round and 
    all methods share the same settings. 
  }
  \label{fig:deepswe-branch-scaling}
\end{figure}

\subsection{Scaling: Depth vs.\ Breadth}
We compare single-round branching with iterative review-and-branch using
GPT-5.6 Luna on DeepSWE (Figure~\ref{fig:deepswe-branch-scaling}).
A configuration $R\times B$ allows up to $R$ review rounds with $B$ branches
per round. All configurations reuse the same baseline rollout for each
task; the baseline resolves 52 tasks, leaving 61 for branching. An analyst
summarizes each newly failed trajectory with its verifier verdict, and a
reviewer combines the accumulated analyses to select $B$ checkpoints and
write continuations. The actor resumes from the corresponding snapshots
and conversation prefixes. Each round executes all $B$ branches, and the
search stops at the end of the first round with a verifier pass.
For single-round scaling, we randomly permute each failed task's 32 accepted
branches once and retain the first $B$ branches for
$B\in\{1,2,4,8,16,32\}$, yielding nested subsets without replacement.
Token totals include the shared
baseline once and the actor, analyst, and reviewer calls attributed to
each setting; single-round subsets include the selected actors and all
recorded usage of the original 32-branch analyst/reviewer once.

\textbf{Depth improves token efficiency.}
Breadth saturates quickly: one branch lifts the baseline from 46.0\% to
68.1\%, but widening from $1\times4$ to $1\times32$ raises the resolve rate
only from 75.2\% to 77.9\% while recorded tokens grow from 0.57B to 1.85B.
Across the comparisons at maximum branch counts of 2, 4, and 8, allocating
more rounds increases the resolve rate and reduces accounted token use.
With an eight-branch maximum, $1\times8$, $2\times4$, $4\times2$, and
$8\times1$ resolve 76.1\%, 85.8\%, 89.4\%, and 91.2\%, using 745.2M,
673.6M, 566.7M, and 525.7M tokens, respectively. The two extremes recover
34 and 51 of the 61 initially failed tasks. Even $2\times1$ reaches
80.5\% with 465.5M tokens, exceeding $1\times16$ at 77.0\% and 1.10B
tokens. These results compare the fixed single-round subsets with the
recorded multi-round runs.

\textbf{Sequential depth limits intra-task parallelism.}
The token-efficiency advantage of depth comes with sequential dependencies:
each review round requires trajectories and verifier feedback from the
preceding round. Branches within a single round can, in principle, execute
concurrently, potentially reducing per-task latency. Intermediate
configurations such as $2\times4$ and $4\times2$ therefore trade off feedback
depth against intra-task parallelism. Our current implementation executes
branches within each round sequentially, so the reported results
characterize the success–token trade-off and do not measure the latency
achievable with parallel branching.
\section{Conclusion}

In this paper, we have proposed \textbf{\textsc{SPROUT}} (\textbf{S}na\textbf{P}shot-based \textbf{R}oll\textbf{OUT}), a framework that addresses the scarcity of verified positive trajectories in long-horizon agent learning by resuming failed rollouts from any intermediate snapshot. Combining copy-on-write filesystem checkpointing, step-aligned conversation restoration, and verifier-guided backtracking, it lets agents backtrack to an earlier state and reuse the trajectory up to that point rather than re-executing it from scratch. Across multiple models and coding benchmarks, \textsc{SPROUT} matches or exceeds repeated independent sampling at a fraction of the token cost, and in RL training it improves the base policy on all benchmarks while outperforming GRPO with shorter training steps. By transforming failed rollouts into cost-effective starting points for exploration, \textsc{SPROUT} provides a practical foundation for collecting positive supervisory signals and scaling autonomous agent capabilities.

\begingroup
\small
\bibliographystyle{plainnat}
\bibliography{references}

@article{dong2024selfcollaboration,
  title     = {Self-Collaboration Code Generation via {ChatGPT}},
  author    = {Dong, Yihong and Jiang, Xue and Jin, Zhi and Li, Ge},
  journal   = {ACM Transactions on Software Engineering and Methodology},
  volume    = {33},
  number    = {7},
  pages     = {1--38},
  year      = {2024},
  publisher = {Association for Computing Machinery (ACM)},
  doi       = {10.1145/3672459},
}

@misc{hsu2026opencollab,
      title={OpenCollab: A Multi-Agent Coding Framework with Programmable Collaboration and Controllable Runtime}, 
      author={Chun-Wah Hsu and Kai Gong and Yu Wu and Xianhe Chen and Mengyang Liu and Jie Li and Hanyu Li and Zhixuan Liu and Naisheng Tang and Jiaying Chi and Ziheng Fan and Xuning He and Xiaokang Yang and Xue Jiang and Yihong Dong},
      year={2026},
      eprint={2609.38345},
      archivePrefix={arXiv},
      primaryClass={cs.SE},
      url={https://arxiv.org/abs/2609.38345}, 
}

@inproceedings{agache2020firecracker,
  title={Firecracker: Lightweight virtualization for serverless applications},
  author={Agache, Alexandru and Brooker, Marc and Iordache, Alexandra and Liguori, Anthony and Neugebauer, Rolf and Piwonka, Phil and Popa, Diana-Maria},
  booktitle={17th USENIX symposium on networked systems design and implementation (NSDI 20)},
  pages={419--434},
  year={2020}
}

@inproceedings{antoniades2025swe,
  title={Swe-search: Enhancing software agents with monte carlo tree search and iterative refinement},
  author={Antoniades, Antonis and {\"O}rwall, Albert and Zhang, Kexun and Xie, Yuxi and Goyal, Anirudh and Wang, William},
  booktitle={International Conference on Learning Representations},
  volume={2025},
  pages={64485--64515},
  year={2025}
}

@article{brown2024large,
  title={Large language monkeys: Scaling inference compute with repeated sampling},
  author={Brown, Bradley and Juravsky, Jordan and Ehrlich, Ryan and Clark, Ronald and Le, Quoc V and R{\'e}, Christopher and Mirhoseini, Azalia},
  journal={arXiv preprint arXiv:2407.21787},
  year={2024}
}

@article{huang2026deepswe,
  title={DeepSWE: Measuring Frontier Coding Agents on Original, Long-Horizon Engineering Tasks},
  author={Huang, Wenqi and Lee, Charley and Tng, Leonard and Ge, Serena},
  journal={arXiv preprint arXiv:2607.07946},
  year={2026}
}

@article{ecoffet2021first,
  title={First return, then explore},
  author={Ecoffet, Adrien and Huizinga, Joost and Lehman, Joel and Stanley, Kenneth O and Clune, Jeff},
  journal={Nature},
  volume={590},
  number={7847},
  pages={580--586},
  year={2021},
  publisher={Nature Publishing Group UK London}
}

@inproceedings{gou2024critic,
  title={Critic: Large language models can self-correct with tool-interactive critiquing},
  author={Gou, Zhibin and Shao, Zhihong and Gong, Yeyun and Yang, Yujiu and Duan, Nan and Chen, Weizhu and others},
  booktitle={International Conference on Learning Representations},
  volume={2024},
  pages={57734--57811},
  year={2024}
}

@inproceedings{hao2023reasoning,
  title={Reasoning with language model is planning with world model},
  author={Hao, Shibo and Gu, Yi and Ma, Haodi and Hong, Joshua and Wang, Zhen and Wang, Daisy and Hu, Zhiting},
  booktitle={Proceedings of the 2023 conference on empirical methods in natural language processing},
  pages={8154--8173},
  year={2023}
}

@article{he2026branching,
  title={Branching Policy Optimization: Sandbox-Native Language Agent Reinforcement Learning},
  author={He, Bowei and Chen, Yankai and Zhang, Xiaokun and Liu, Xue},
  journal={arXiv preprint arXiv:2607.14171},
  year={2026}
}

@inproceedings{hou2025treerl,
  title={Treerl: Llm reinforcement learning with on-policy tree search},
  author={Hou, Zhenyu and Hu, Ziniu and Li, Yujiang and Lu, Rui and Tang, Jie and Dong, Yuxiao},
  booktitle={Proceedings of the 63rd Annual Meeting of the Association for Computational Linguistics (Volume 1: Long Papers)},
  pages={12355--12369},
  year={2025}
}

@article{kazemnejad2024vineppo,
  title={Vineppo: Refining credit assignment in rl training of llms},
  author={Kazemnejad, Amirhossein and Aghajohari, Milad and Portelance, Eva and Sordoni, Alessandro and Reddy, Siva and Courville, Aaron and Roux, Nicolas Le},
  journal={arXiv preprint arXiv:2410.01679},
  year={2024}
}

@article{li2025agentgit,
  title={AgentGit: A version control framework for reliable and scalable LLM-powered multi-agent systems},
  author={Li, Yang and Ping, Siqi and Chen, Xiyu and Qi, Xiaojian and Wang, Zigan and Luo, Ye and Zhang, Xiaowei},
  journal={arXiv preprint arXiv:2511.00628},
  year={2025}
}

@article{merkel2014docker,
  title={Docker: lightweight linux containers for consistent development and deployment},
  author={Merkel, Dirk and others},
  journal={Linux j},
  volume={239},
  number={2},
  pages={2},
  year={2014}
}

@article{shao2024deepseekmath,
  title={Deepseekmath: Pushing the limits of mathematical reasoning in open language models},
  author={Shao, Zhihong and Wang, Peiyi and Zhu, Qihao and Xu, Runxin and Song, Junxiao and Bi, Xiao and Zhang, Haowei and Zhang, Mingchuan and Li, YK and Wu, Yang and others},
  journal={arXiv preprint arXiv:2402.03300},
  year={2024}
}

@article{shinn2023reflexion,
  title={Reflexion: Language agents with verbal reinforcement learning},
  author={Shinn, Noah and Cassano, Federico and Gopinath, Ashwin and Narasimhan, Karthik and Yao, Shunyu},
  journal={Advances in neural information processing systems},
  volume={36},
  pages={8634--8652},
  year={2023}
}

@article{wang2024executable,
  title={Executable code actions elicit better llm agents},
  author={Wang, Xingyao and Chen, Yangyi and Yuan, Lifan and Zhang, Yizhe and Li, Yunzhu and Peng, Hao and Ji, Heng},
  journal={arXiv preprint arXiv:2402.01030},
  year={2024}
}

@article{wang2026fork,
  title={Fork, Explore, Commit: OS Primitives for Agentic Exploration},
  author={Wang, Cong and Zheng, Yusheng},
  journal={arXiv preprint arXiv:2602.08199},
  year={2026}
}

@article{wang2026contrastive,
  title={Contrastive Branch Policy Optimization},
  author={Wang, Ying and Qiu, Changlin and Lin, Bang and Jin, Linbo and Jiang, Wen and Sun, Zhe and Yang, Jingli},
  journal={arXiv preprint arXiv:2608.24300},
  year={2026}
}

@inproceedings{wu2026spark,
  title={Spark: Strategic policy-aware exploration via dynamic branching for long-horizon agentic learning},
  author={Wu, Jinyang and Yang, Shuo and Shen, Yuhao and Zhang, Shuai and Wen, Zhengqi and Tao, Jianhua},
  booktitle={Proceedings of the 64th Annual Meeting of the Association for Computational Linguistics (Volume 1: Long Papers)},
  pages={23981--24004},
  year={2026}
}

@article{yang2024swe,
  title={Swe-agent: Agent-computer interfaces enable automated software engineering},
  author={Yang, John and Jimenez, Carlos and Wettig, Alexander and Lieret, Kilian and Yao, Shunyu and Narasimhan, Karthik and Press, Ofir},
  journal={Advances in Neural Information Processing Systems},
  volume={37},
  pages={50528--50652},
  year={2024}
}

@article{yao2022react,
  title={React: Synergizing reasoning and acting in language models},
  author={Yao, Shunyu and Zhao, Jeffrey and Yu, Dian and Du, Nan and Shafran, Izhak and Narasimhan, Karthik and Cao, Yuan},
  journal={arXiv preprint arXiv:2210.03629},
  year={2022}
}

@article{yao2023tree,
  title={Tree of thoughts: Deliberate problem solving with large language models},
  author={Yao, Shunyu and Yu, Dian and Zhao, Jeffrey and Shafran, Izhak and Griffiths, Tom and Cao, Yuan and Narasimhan, Karthik},
  journal={Advances in neural information processing systems},
  volume={36},
  pages={11809--11822},
  year={2023}
}

@article{zhou2023language,
  title={Language agent tree search unifies reasoning acting and planning in language models},
  author={Zhou, Andy and Yan, Kai and Shlapentokh-Rothman, Michal and Wang, Haohan and Wang, Yu-Xiong},
  journal={arXiv preprint arXiv:2310.04406},
  year={2023}
}

@inproceedings{wang2025openhands,
  title={Openhands: An open platform for ai software developers as generalist agents},
  author={Wang, Xingyao and Li, Boxuan and Song, Yufan and Xu, Frank F and Tang, Xiangru and Zhuge, Mingchen and Pan, Jiayi and Song, Yueqi and Li, Bowen and Singh, Jaskirat and others},
  booktitle={International Conference on Learning Representations},
  volume={2025},
  pages={65882--65919},
  year={2025}
}

@article{xu2025toward,
  title={Toward Systems Foundations for Agentic Exploration},
  author={Xu, Jiakai and Zhou, Tianle and Wu, Eugene and Kaffes, Kostis},
  journal={arXiv preprint arXiv:2510.05556},
  year={2025}
}

@inproceedings{weisz2021query,
  title={On query-efficient planning in mdps under linear realizability of the optimal state-value function},
  author={Weisz, Gellert and Amortila, Philip and Janzer, Barnab{\'a}s and Abbasi-Yadkori, Yasin and Jiang, Nan and Szepesv{\'a}ri, Csaba},
  booktitle={Conference on Learning Theory},
  pages={4355--4385},
  year={2021},
  organization={PMLR}
}

@inproceedings{yin2022efficient,
  title={Efficient local planning with linear function approximation},
  author={Yin, Dong and Hao, Botao and Abbasi-Yadkori, Yasin and Lazi{\'c}, Nevena and Szepesv{\'a}ri, Csaba},
  booktitle={International Conference on Algorithmic Learning Theory},
  pages={1165--1192},
  year={2022},
  organization={PMLR}
}

@article{brown2024monkeys,
  author    = {Brown, Bradley and Juravsky, Jordan and Ehrlich, Ryan and Clark, Ronald and Le, Quoc V. and R{\'e}, Christopher and Mirhoseini, Azalia},
  title     = {Large Language Monkeys: Scaling Inference Compute with Repeated Sampling},
  journal   = {arXiv preprint arXiv:2407.21787},
  year      = {2024}
}

@article{datacurve2026deepswe,
  author    = {Huang, Wenqi and Lee, Charley and Tng, Leonard and Ge, Serena},
  title     = {{DeepSWE}: Measuring Frontier Coding Agents on Original, Long-Horizon Engineering Tasks},
  journal   = {arXiv preprint arXiv:2607.07946},
  year      = {2026}
}

@article{he2026bpo,
  author    = {He, Bowei and Chen, Yankai and Zhang, Xiaokun and Liu, Xue},
  title     = {Branching Policy Optimization: Sandbox-Native Language Agent Reinforcement Learning},
  journal   = {arXiv preprint arXiv:2607.14171},
  year      = {2026}
}

@inproceedings{jimenez2024swebench,
  author    = {Jimenez, Carlos E. and Yang, John and Wettig, Alexander and Yao, Shunyu and Pei, Kexin and Press, Ofir and Narasimhan, Karthik},
  title     = {{SWE}-bench: Can Language Models Resolve Real-World {GitHub} Issues?},
  booktitle = {International Conference on Learning Representations (ICLR)},
  year      = {2024}
}

@article{li2022alphacode,
  author    = {Li, Yujia and Choi, David and Chung, Junyoung and Kushman, Nate and Schrittwieser, Julian and Leblond, R{\'e}mi and Eccles, Tom and Keeling, James and Gimeno, Felix and Dal Lago, Agustin and others},
  title     = {Competition-level code generation with {AlphaCode}},
  journal   = {Science},
  volume    = {378},
  number    = {6624},
  pages     = {1092--1097},
  year      = {2022}
}

@article{shao2024grpo,
  author    = {Shao, Zhihong and Wang, Peiyi and Zhu, Qihao and Xu, Runxin and Song, Junxiao and Bi, Xiao and Zhang, Haowei and Zhang, Mingchuan and Li, Y. K. and Wu, Yang and others},
  title     = {{DeepSeekMath}: Pushing the Limits of Mathematical Reasoning in Open Language Models},
  journal   = {arXiv preprint arXiv:2402.03300},
  year      = {2024}
}

@article{wang2026cbpo,
  author    = {Wang, Ying and Qiu, Changlin and Lin, Bang and Jin, Linbo and Jiang, Wen and Sun, Zhe and Yang, Jingli},
  title     = {Contrastive Branch Policy Optimization},
  journal   = {arXiv preprint arXiv:2608.24300},
  year      = {2026}
}

@inproceedings{yang2024sweagent,
  author    = {Yang, John and Jimenez, Carlos E. and Wettig, Alexander and Lieret, Kilian and Yao, Shunyu and Narasimhan, Karthik and Press, Ofir},
  title     = {{SWE}-agent: Agent-Computer Interfaces Enable Automated Software Engineering},
  booktitle = {Advances in Neural Information Processing Systems (NeurIPS)},
  volume    = {37},
  pages     = {50528--50652},
  year      = {2024}
}

@inproceedings{yao2023tot,
  author    = {Yao, Shunyu and Yu, Dian and Zhao, Jeffrey and Shafran, Izhak and Griffiths, Thomas L. and Cao, Yuan and Narasimhan, Karthik},
  title     = {Tree of Thoughts: Deliberate Problem Solving with Large Language Models},
  booktitle = {Advances in Neural Information Processing Systems (NeurIPS)},
  volume    = {36},
  pages     = {11809--11822},
  year      = {2023}
}

@inproceedings{zhou2023lats,
  author    = {Zhou, Andy and Yan, Kai and Shlapentokh-Rothman, Michal and Wang, Haohan and Wang, Yu-Xiong},
  title     = {Language Agent Tree Search Unifies Reasoning, Acting, and Planning in Language Models},
  booktitle = {International Conference on Machine Learning (ICML)},
  year      = {2024}
}

@misc{openai2024swebenchverified,
  author       = {{OpenAI}},
  title        = {Introducing {SWE}-bench {V}erified},
  howpublished = {\url{https://openai.com/index/introducing-swe-bench-verified/}},
  year         = {2024},
  note         = {Accessed: 2024}
}

@misc{anthropic2025claudecode,
  author       = {{Anthropic}},
  title        = {Claude Code: An Agentic Coding Tool},
  howpublished = {\url{https://docs.anthropic.com/en/docs/agents-and-tools/claude-code}},
  year         = {2025},
  note         = {Accessed: 2025}
}

@article{deng2025swe,
  title={Swe-bench pro: Can ai agents solve long-horizon software engineering tasks?},
  author={Deng, Xiang and Da, Jeff and Pan, Edwin and He, Yannis Yiming and Ide, Charles and Garg, Kanak and Lauffer, Niklas and Park, Andrew and Pasari, Nitin and Rane, Chetan and others},
  journal={arXiv preprint arXiv:2509.16941},
  year={2025}
}

@article{kumar2026tvcache,
  title={TVCACHE: A stateful tool-value cache for post-training LLM agents},
  author={Kumar, Abhishek Vijaya and Kataria, Bhaskar and Oh, Byungsoo and Manzoor, Emaad and Singh, Rachee},
  journal={arXiv preprint arXiv:2602.10986},
  year={2026}
}

@article{DBLP:journals/corr/abs-2607-24653,
  author       = {Kimi Team},
  title        = {Kimi {K3:} Open Frontier Intelligence},
  journal      = {CoRR},
  volume       = {abs/2607.24653},
  year         = {2026},
  url          = {https://doi.org/10.48550/arXiv.2607.24653},
  doi          = {10.48550/ARXIV.2607.24653},
  eprinttype   = {arXiv},
  eprint       = {2607.24653},
  bibsource    = {dblp computer science bibliography, https://dblp.org}
}

@inproceedings{merrill2026terminal,
  title={Terminal-bench: Benchmarking agents on hard, realistic tasks in command line interfaces},
  author={Merrill, Mike and Shaw, Alexander and Carlini, Nicholas and Li, Boxuan and Raj, Harsh and Bercovich, Ivan and Shi, Lin and Shin, Jeong and Walshe, Thomas and Buchanan, E Kelly and others},
  booktitle={International Conference on Learning Representations},
  volume={2026},
  pages={40903--40986},
  year={2026}
}

@article{yu2026shepherd,
  title={Shepherd: Enabling Programmable Meta-Agents via Reversible Agentic Execution Traces},
  author={Yu, Simon and Chong, Derek and Nandi, Ananjan and Soylu, Dilara and Sun, Jiuding and Manning, Christopher D and Shi, Weiyan},
  journal={arXiv preprint arXiv:2605.10913},
  year={2026}
}

@misc{qwen2026qwen38,
  title={{Qwen3.8-Max}: A New Bar for Coding and Cowork},
  author={{Qwen Team}},
  year={2026},
  month={August},
  howpublished={\url{https://qwen.ai/blog?id=qwen3.8}}
}

@misc{anthropic2026sonnet5,
  title={Introducing Claude Sonnet 5},
  author={{Anthropic}},
  year={2026},
  month={June},
  howpublished={\url{https://www.anthropic.com/news/claude-sonnet-5}}
}

@misc{openai2026gpt56,
  title={GPT-5.6 Luna},
  author={{OpenAI}},
  year={2026},
  howpublished={\url{https://developers.openai.com/api/docs/models/gpt-5.6-luna}}
}

@inproceedings{RL-PLUS,
  author       = {Yihong Dong and
                  Xue Jiang and
                  Yongding Tao and
                  Huanyu Liu and
                  Kechi Zhang and
                  Lili Mou and
                  Rongyu Cao and
                  Yingwei Ma and
                  Jue Chen and
                  Binhua Li and
                  Zhi Jin and
                  Fei Huang and
                  Yongbin Li and
                  Ge Li},
  title        = {{RL-PLUS:} Countering Capability Boundary Collapse of LLMs in Reinforcement
                  Learning with Hybrid-policy Optimization},
  booktitle    = {{ACL} {(1)}},
  pages        = {43029--43053},
  publisher    = {Association for Computational Linguistics},
  year         = {2026}
}

@inproceedings{ROCODE,
  author       = {Xue Jiang and
                  Yihong Dong and
                  Yongding Tao and
                  Huanyu Liu and
                  Zhi Jin and
                  Ge Li},
  title        = {{ROCODE:} Integrating Backtracking Mechanism and Program Analysis
                  in Large Language Models for Code Generation},
  booktitle    = {{ICSE}},
  pages        = {334--346},
  publisher    = {{IEEE}},
  year         = {2025}
}

@misc{deepseek2026v41flash,
  author       = {{DeepSeek-AI}},
  title        = {DeepSeek-V4.1-Flash Technical Report},
  year         = {2026},
  howpublished = {\url{https://huggingface.co/deepseek-ai/DeepSeek-V4.1-Flash/blob/main/DeepSeek_V41_Tech_Report.pdf}},
  note         = {Technical report}
}
\endgroup

\clearpage
\appendix

\section{Experimental Setup and Supplementary Experiments}
\label{app:experiment}

\paragraph{Benchmarks and models.}
We use the 113-task public DeepSWE, the 89-task Terminal-Bench 2.1,
and the 51-task SWE-bench Pro v2 Hard. Sampling experiments use Kimi K3,
Qwen3.8-27B, Claude Sonnet 5, DeepSeek V4.1 Flash, and GPT 5.6 Luna with
fixed model weights. Kimi K3 runs with reasoning effort set to \texttt{max},
and all other models run with reasoning effort set to \texttt{high}. In
\textsc{SPROUT}, the actor also performs failure analysis and branch
selection under separate role prompts, with the same reasoning effort as
its actor role.

\paragraph{Execution environment.}
All runs use mini-swe-agent \citep{yang2024sweagent} with a
sandbox shell as the environment interface. Shell calls are
synchronous, and background processes are disabled. SPROUT
checkpoints the full filesystem after each turn and pairs
it with the corresponding agent context. Each branch resumes
in an isolated sandbox. DeepSWE attempts run offline with 2
CPUs, 8~GB of memory, and a time limit of 10,800 seconds.

\paragraph{Baselines.}
Independent sampling performs four fresh attempts per task.
Pass@1 is the mean resolve rate across the four passes, and
Pass@4 is the fraction of tasks solved in at least one pass.
We also evaluate Shepherd \citep{yu2026shepherd} with its own
fork-selection procedure and an entropy-based variant, which selects branch points
using policy entropy and is only evaluated with Qwen3.8-27B that can expose token log-probabilities in our setup.

\paragraph{Collection protocol.}
For each model and benchmark, a single-attempt pass over all tasks provides the
parent attempts. \textsc{SPROUT} branches from the failed parents, restoring
each selected checkpoint together with its step-aligned agent context, and a
task counts as solved if its parent or any of its branches passes the
verifier.

\paragraph{Reported metrics.}
The sampling comparison reports resolve rate, recovery rate,
total execution steps, and total tokens. The resolve rate counts
tasks with at least one verified success. Token totals combine
input and output tokens and are expressed in millions.We report normalized recovery as \((R_{\text{method}}-R_{\text{Pass@1}})/(1-R_{\text{Pass@1}})\), measuring the fraction of the remaining performance gap beyond Pass@1 recovered by each method.

\begin{table}[h]
\centering
\small
\caption{RL hyperparameters and training configuration.}
\label{tab:rl-training-config}
\begin{tabular}{ll}
\toprule
\textbf{Parameter} & \textbf{Value} \\
\midrule
Model & Qwen3.8-27B \\
Fine-tuning & Full parameters \\
Prompts per update & 32 \\
Rollouts per prompt & 8 \\
Nominal global batch size & 256 rollouts \\
Maximum context length & 128K tokens \\
Sampling temperature & 1.0 \\
Thinking mode & Enabled \\
Reasoning effort & \texttt{xhigh} \\
Optimizer & Adam \\
Learning rate & $2\times10^{-6}$ \\
Learning-rate schedule & Constant \\
KL penalty coefficient & 0 \\
Hardware & $8\times$ NVIDIA B300 \\
Tensor parallelism & 2 \\
Context parallelism & 4 \\
Microbatching & Dynamic \\
Microbatch token cap per GPU & 32,768 \\
\bottomrule
\end{tabular}
\end{table}

\paragraph{RL training.}
We fine-tune Qwen3.8-27B on SWE-bench Rebench v2 Filtered and select the top 2,115 tasks ranked by golden patch
length and construct groups of eight evaluated attempts, comprising two
independent roots and three children per root. The rollout orchestration
is detailed in Appendix~\ref{app:algorithm}.
Basic GRPO uses the same training set and samples eight
independent rollouts per prompt. SPROUT RL and basic GRPO share the same training configuration, detailed in Table~\ref{tab:rl-training-config}, and report the score under the official evaluation protocol.

\paragraph{Additional evaluation with OpenCollab.} We additionally evaluate Qwen3.8-27B using OpenCollab (Base) \citep{hsu2026opencollab} on the DeepSWE benchmark, following the experimental settings of Table~\ref{tab:sampling-matrix}. \textsc{SPROUT} achieves resolve rates of 73.45\% with 112.8\% relative improvement compared with basic rollout.
\begin{algorithm}[h]
\caption{\textsc{SPROUT} training iteration for one task $x$}
\label{alg:sprout}
\begin{algorithmic}[1]
\Require policy $\pi_\theta$ (also acting as analyst, reviewer, and turn
 writer), verifier $\mathcal{V}_x$, judge $J$, family size $K$ ($=3$),
 oversampling $O$ ($=1$), context budget $L$, headroom $\Delta$
\State Sample two independent rollouts $\tau_1,\tau_2\sim\rho_\theta(\cdot\mid s_0)$,
 checkpointing every turn boundary; evaluate each with $\mathcal{V}_x$
\State $\mathcal{A}_x\leftarrow\{\tau_1,\tau_2\}$;\quad
 $m\leftarrow\textsc{TT}$ if $R_x(\tau_1)=R_x(\tau_2)=1$,
 $\textsc{FF}$ if $R_x(\tau_1)=R_x(\tau_2)=0$, else $\textsc{TF}$
\For{each root $i\in\{1,2\}$}
 \State $\mathcal{B}_i\leftarrow\{t\in\mathcal{E}(\tau_i) : t\le t^{\mathrm{cut}}_i,\
  |\tau_{i,\le t}|+\Delta\le L\}$
  \Comment{usable boundaries}
 \If{$m\in\{\textsc{TT},\textsc{TF}\}$}
  \State $\hat t_i\leftarrow$ boundary in $\mathcal{B}_i$ selected by the reviewer
  \State $(\hat t_{i,j},a'_{i,j})\leftarrow(\hat t_i,\varnothing)$ for $j=1,\ldots,K{+}O$
   \Comment{unguided continuation}
 \Else
  \Comment{$m=\textsc{FF}$: guided repair}
  \State $\{(t_k,g_k)\}_k\leftarrow$ ranked boundaries $t_k\in\mathcal{B}_i$ with
   corrective directions $g_k$, from a review of $\mathcal{A}_x$ including verifier feedback
  \State $a'_k\leftarrow$ starting turn written from $\tau_{i,\le t_k}$ and $g_k$ only;
   keep $(t_k,a'_k)$ if accepted by $J$
  \State $(\hat t_{i,j},a'_{i,j})\leftarrow(t_j,a'_j)$ for $j=1,\ldots,K$;\quad
   $(\hat t_{i,K+o},a'_{i,K+o})\leftarrow(t_o,a'_o)$ for $o=1,\ldots,O$
 \EndIf
 \For{$j=1,\ldots,K{+}O$ \textbf{in parallel}}
  \State $e'_j\leftarrow\operatorname{Restore}(c_{i,\hat t_{i,j}})$,\;
   $h'_j\leftarrow\operatorname{Recover}(c_{i,\hat t_{i,j}})$ in a fresh sandbox
  \If{$a'_{i,j}\neq\varnothing$}
   \State Execute the tool calls of $a'_{i,j}$ in order to obtain $o'_j$;\;
    $s_j^+\leftarrow$ joint state after the context update
  \Else
   \State $s_j^+\leftarrow(e'_j,h'_j)$
  \EndIf
  \State Sample $\eta_j\sim\rho_\theta(\cdot\mid s_j^+)$;\;
   $\tau'_{i,j}\leftarrow\tau_{i,\le\hat t_{i,j}}\oplus(a'_{i,j},o'_j)\oplus\eta_j$
   \Comment{$(a'_{i,j},o'_j)$ omitted if $a'_{i,j}=\varnothing$}
  \State Evaluate $\tau'_{i,j}$ with $\mathcal{V}_x$ and add it to $\mathcal{A}_x$
 \EndFor
 \State Retain the first $K$ children of family $i$ that complete evaluation; cancel the rest
\EndFor
\If{a family has fewer than $K$ completed children}
 \State Discard $x$ and draw a replacement task
\EndIf
\State $\mathcal{T}_x\leftarrow\{\tau_1,\tau_2\}\cup\{\tau'_{i,j}\}_{\text{retained}}$,
 $|\mathcal{T}_x|=8$; compute advantages and loss masks
 (Equation~\ref{eq:tree-credit})
\State Update $\pi_\theta$ with one policy-gradient step over the batch of all
 tasks (Equation~\ref{eq:tree-loss})
\end{algorithmic}
\end{algorithm}

\section{Training Algorithm}
\label{app:algorithm}

Algorithm~\ref{alg:sprout} details one \textsc{SPROUT} training iteration for
a single task $x$. Two independent roots are first sampled from $s_0$ and
graded, and their outcomes determine the mode of the group: TT if both
succeed, FF if both fail, and TF otherwise. Each root $\tau_i$ then grows a
family of $K$ children from its usable boundaries $\mathcal{B}_i$, namely the
eligible boundaries (Section~\ref{sec:restorable-states}) whose prefix leaves
at least $\Delta$ tokens of headroom within the context budget $L$. In TT and TF groups, the reviewer selects
one boundary per root, and all children of that root continue from it without
an injected starting turn. In FF groups, the reviewer analyzes the archive,
including verifier feedback, and returns ranked boundaries $t_k$, each paired
with a corrective direction $g_k$. The policy then writes a starting turn
$a'_k$ at each boundary from the visible prefix and $g_k$ alone, without
access to the analysis or the verifier feedback, and a judge filters the
resulting turns. Each child restores the environment and context at its
boundary in a fresh sandbox, executes its starting turn if it has one,
continues under the current policy, and is graded. Each family launches $O$
extra children so that stragglers do not delay the iteration, and only the
first $K$ children to complete evaluation are retained. A task with a family
that has fewer than $K$ completed children is replaced. The eight retained attempts are credited and
masked as in Equation~\ref{eq:tree-credit} and contribute to one
policy-gradient step (Equation~\ref{eq:tree-loss}).
\section{Starting Turns versus User Hints}
\label{app:hint-format}

A guided branch needs some way to carry the corrective direction into the
restored context. We considered two formats.

\paragraph{User hints.}
The simplest format is a \emph{user hint}: an external message, appended to
the restored context, that states the corrective direction. Because the hint
is derived from verifier feedback, it tends to leak into the continuation.
The actor often recites the message, quotes test names or expected values
from it, or justifies later actions by referring to it, so the continuation
depends on information that is unavailable during ordinary execution. The
hint also causes a problem for training. It is not an action of the policy
and must be removed from the learner's context. Once it is removed, the
training context no longer explains the continuation: turns that recite or
respond to the hint remain in the trajectory without a visible cause.

\paragraph{Starting turns.}
\textsc{SPROUT} instead writes the guidance as a \emph{starting turn}
$a'_j$ in the actor's own voice, consisting of brief reasoning followed by
tool calls whose observations come from real execution
(Section~\ref{sec:method}). The turn must read as a natural continuation of
the prefix and may not mention checkpoints, other branches, or hidden tests,
and in training groups a judge filters turns that violate these constraints
(Appendix~\ref{app:algorithm}). The context therefore contains no foreign
message. During training, $a'_j$ remains in place as an ordinary policy turn,
and only the review context used to write it is removed
(Section~\ref{sec:training}). The actor thus learns the corrective action
itself rather than how to respond to an external hint.

\section{State Projection and Cache Matching Efficiency}
\label{app:state-matching}

\paragraph{Motivation.}
In agentic reinforcement learning for long-horizon coding tasks, rollouts frequently traverse identical or semantically equivalent intermediate states (e.g., repository navigation, virtual environment configuration, dependency installation). 
Re-executing these initial setup steps from scratch wastes compute budget, while naively snapshotting and restoring complete microVM environments can introduce severe I/O contention. 
To amortize rollout overhead, \textsc{SPROUT} incorporates a state-matching mechanism that identifies whether an intermediate execution state can be directly resumed from an existing checkpoint. 
This appendix provides a detailed empirical comparison of our state projection approach against representative caching and hashing baselines.

\paragraph{Experimental Protocol and Baselines.}
We evaluate caching mechanisms on 248 rollout trajectories (136 generated with GPT 5.6 Luna and 112 with Qwen as query policies) collected across the 134 candidate tasks that initially failed single-pass exploration. 
We evaluate five configurations:
\begin{itemize}
    \item \textbf{No Cache}: The standard baseline that disables state reuse entirely, executing every trajectory prefix from scratch.
    \item \textbf{Exact Full History}: Matches intermediate execution states strictly by verifying identical raw conversational histories and tool-call logs.
    \item \textbf{Official TVCACHE TCG}: The official Tool Call Graph (TCG) prefix-matching implementation from TVCACHE~\citep{kumar2026tvcache}, a stateful execution cache designed for agent post-training.
    \item \textbf{Full-Workspace Hash (Upper Bound)}: Computes recursive SHA-256 cryptographic hashes across the entire microVM workspace filesystem after each tool step to detect identical on-disk states. This represents the empirical upper bound of state-level hit rates.
    \item \textbf{Ours (Scan-free Relaxed Projection)}: Projects filesystem mutations into lightweight metadata signatures at tool boundaries without recursive disk-tree traversal, filtering out superficial non-deterministic side effects.
\end{itemize}

\begin{table}[t]
\centering
\caption{\textbf{Evaluation of intermediate state matching and caching mechanisms.} Evaluated on 248 trajectories across 134 candidate tasks. Hit rates report both the total ratio and the breakdown by query policy. Online latency measures the verification time per candidate checkpoint, and retained candidates denote unresolved task slots remaining after cache reuse.}
\label{tab:state-matching}
\resizebox{\linewidth}{!}{%
\setlength{\tabcolsep}{4pt}
\begin{tabular}{llcccccc}
\toprule
\textbf{Method} & \textbf{Source / Setting} & \textbf{Total Hit Rate} & \textbf{Luna} & \textbf{Qwen} & \textbf{Precision} & \textbf{Latency} & \textbf{Retained} \\
\midrule
No Cache & Baseline & 0/248 & 0/136 & 0/112 & --- & 0\,ms & 134/134 \\
Exact Full History & Strict Trace Match & 2/248 (0.8\%) & 2/136 (1.5\%) & 0/112 (0.0\%) & 100\% & 0.0003\,ms & 133/134 \\
Official TVCACHE TCG & TVCACHE~\citep{kumar2026tvcache} & 8/248 (3.2\%) & 8/136 (5.9\%) & 0/112 (0.0\%) & 100\% & 0.0017\,ms & 133/134 \\
Full-Workspace Hash & Exhaustive Scan & 91/248 (36.7\%) & 65/136 (47.8\%) & 26/112 (23.2\%) & 100\% & 79.1\,ms & 38/134 \\
\midrule
\textbf{Ours} & \textbf{Relaxed Projection} & \textbf{35/248 (14.1\%)} & \textbf{28/136 (20.6\%)} & \textbf{7/112 (6.2\%)} & \textbf{100\%} & \textbf{0.201\,ms} & \textbf{82/134} \\
\bottomrule
\end{tabular}%
}
\end{table}

\paragraph{Analysis and Key Takeaways.}
Table~\ref{tab:state-matching} highlights the latency-coverage trade-offs across different state-matching paradigms:
\begin{enumerate}
    \item \textbf{Brittleness of Textual and Graph Matching}: 
    Exact full history matching yields an extremely low hit rate ($0.8\%$) because trivial non-deterministic variations in tool outputs (e.g., timestamps, process IDs, temporary directory names) invalidate string identity. Similarly, while TVCACHE's TCG tracks tool call trees, it misses many reusable states ($3.2\%$ hit rate) due to subtle command-line syntax variations in long bash command sequences.
    
    \item \textbf{The Overhead Bottleneck of Deep Hashing}: 
    Full-workspace hashing achieves the theoretical upper bound hit rate of $36.7\%$ (reducing retained tasks from 134 to 38). However, traversing and hashing multi-gigabyte repository workspaces incurs an average online latency of $79.1$\,ms per checkpoint check. In high-throughput parallel rollout environments where hundreds of microVMs execute synchronously, this overhead causes severe CPU and disk I/O contention.
    
    \item \textbf{Optimal Latency--Coverage Balance}: 
    Our \emph{Scan-free Relaxed Projection} strikes a practical balance: it achieves a $14.1\%$ overall hit rate ($20.6\%$ on Luna queries) while requiring only $0.201$\,ms of online latency---over $390\times$ faster than full-workspace hashing. Crucially, all matched states achieve $100\%$ precision, ensuring that branched continuations never inherit corrupted or mismatched filesystem states.
    
    \item \textbf{Model-Dependent Trajectory Divergence}: 
    Across all caching methods, trajectories generated by GPT 5.6 Luna exhibit significantly higher hit rates than Qwen (e.g., $20.6\%$ vs.\ $6.2\%$ under our method, and $47.8\%$ vs.\ $23.2\%$ under workspace hashing). Luna follows more standardized repository inspection patterns, producing reusable environment states during early setup stages.
\end{enumerate}

\tcbset{promptbox/.style={
  enhanced, breakable, listing only,
  colback=LabPanel, colframe=LabRule, coltitle=LabAccent, colbacktitle=LabPanel,
  fonttitle=\sffamily\bfseries\small, boxrule=0.5pt, arc=1.4mm, left=4pt, right=4pt, top=3pt, bottom=3pt,
  listing options={basicstyle=\ttfamily\footnotesize, breaklines=true, breakatwhitespace=false,
                   columns=fullflexible, keepspaces=true, showstringspaces=false,
                   postbreak=\mbox{\textcolor{gray}{$\hookrightarrow$}\space}}}}

\section{Prompts of the branching controller}
\label{app:prompts}

The controller runs an \emph{analyst} once per failed attempt and a \emph{reviewer} once per
round. The reviewer's plan is validated before any branch executes; a rejected plan is returned
to the reviewer with the validator's message (at most three attempts). In the assistant-turn
mode used in our experiments the reviewer authors the next assistant message of each branch
(reasoning plus \texttt{bash} tool calls), which is spliced into the retained conversation
prefix; the actor therefore never receives a user-side hint. Placeholders in braces
(\texttt{\{problem\}}, \texttt{\{transcript\}}, \ldots) are filled by the controller.

\subsection{Analyst prompt (one per failed attempt)}
\begin{tcblisting}{promptbox, title={Analyst prompt}, label={prompt:analyst}}
You are analysing ONE failed coding attempt. Your private analysis will be
pooled with other attempts so a reviewer can choose a checkpoint and useful
directions for continuing from it. The continuing agent remembers only the
conversation up to that checkpoint.

The goal is a natural and effective continuation: if the added direction is
later removed, the next reasoning and actions should still make sense from the
retained work and the agent's own observations. Diagnose the missed code-level
connection that could plausibly occur to the agent at that point.

## The problem it was solving
{problem}

## Private grading evidence (not observed by the continuing agent)
{verdict}

## The attempt, one line per tool step ("[N] tool(args) -> result")
{transcript}

## Recorded snapshot/session pairs available for selection
{checkpoint_steps}

## Translate evidence into repair directions
- Your output should help the reviewer choose a repair direction, not reproduce
  the grading report. In every returned field, omit verifier-only filenames,
  test names/IDs and grader paths. Translate their useful information into the
  affected program behavior, code relationship or invariant instead.
- A path in verifier output is not evidence that the actor can open it. Refer
  to a repository file or symbol only when the task or retained prefix
  establishes it is available at that candidate checkpoint. Do not ask the
  actor to read, locate or recreate a hidden test, or retrieve a grading artifact.
  Suggest inspecting accessible implementation/callers or constructing a small
  local check from the public task and available code instead.
- If the cause is established, state the correction directly and explain the
  relevant constraint. Otherwise give the strongest supported hypothesis: what
  code relationship may be involved, why it is plausible, what remains unknown,
  and one accessible check with what would support or rule it out. If no specific
  cause is supported, say the cause is unresolved and identify the nearest
  evidence-backed area to investigate, rather than inventing a precise fix.
- Do not merely add 'possibly' to an invented explanation. In particular, do
  not guess how an unseen test constructs its fixtures, router groups or inputs.
  A symptom does not establish those details. Prefer a check that distinguishes
  plausible mechanisms before recommending a behavior-changing edit.
- Keep private observations distinct from what the actor already knows. In
  lesson and each candidate's why, express a code-level direction and its
  connection to the retained work, not instructions to consult external evidence.

Example of translating a route-lookup failure into a qualified direction:
"The route prefix may be composed inconsistently between the registration helper
and its caller. Trace the group prefix and route suffix in the available code,
then compare the resulting URL with the task's required endpoint. Check for a
missing or duplicated segment before changing registration. The evidence does
not yet establish whether a caller-owned or handler-owned prefix is appropriate."
This is an example of reasoning, not a diagnosis to reuse for unrelated tasks.

## Evidence and output rules
- Use the verdict and the whole transcript as private diagnostic evidence.
  failure_reason and lesson are controller-only reports, not actor messages.
  Retain the specific mechanism, useful correction and important constraints;
  do not dilute a diagnosis into generic advice to inspect code or run tests.
- Separate observed facts from inferences. Distinguish the required behavior,
  a suspected cause, and a possible repair. An external failure can motivate
  a hypothesis without proving its cause. Do not change a behavioral contract
  simply because another implementation would be easier to check.
- For each candidate, identify the code, decision or observation that connects
  the retained work to the next useful question. Explain what a concrete action
  would establish, without assuming the agent has seen the discarded suffix.
- Preserve behavior outside the suspected bug. Identify relevant accepted
  inputs, boundary cases or API responsibilities that a repair must retain.
  Express the issue in program behavior rather than relying on test labels.
- Do not compose the continuing agent's inner monologue or invent observations.
  The reviewer needs a useful lead, not text the agent must repeat or a story
  about receiving feedback.

## Checkpoint rules
- Tool steps span {lo}..{hi}, but only the recorded steps listed above have
  eligible snapshot/session pairs. Choose candidates from that list; a matching
  native conversation cut is checked separately before launch. A branch at
  step N must resume the environment AS IT STOOD AFTER step N.
- branch_candidates: up to {candidate_limit} useful steps, not a quota. LATER IS BETTER when it preserves
  sound work, but choose before the decisive wrong turn. If the approach was
  already wrong at the start, propose the earliest eligible point and explain why.
- Describe salvage relative to the state after each candidate step, not the
  final tree. Later edits, files, dependencies and observations are not already
  present at an earlier checkpoint.
- Each candidate's why should name the prefix connection and the next useful
  check. Avoid requiring the continuation to redo reasoning or checks that
  the retained prefix already established.

Return ONLY a JSON object, no prose:
{"failure_reason": "<observed behavior; supported mechanism or hypothesis; remaining uncertainty>",
  "lesson": "<concrete repair direction or discriminating check; constraints to preserve>",
  "salvage": "<work present at the candidate steps, or 'nothing'>",
  "branch_candidates": [{"step": <int>, "why": "<visible prefix connection; next check and what it distinguishes>"}]}
\end{tcblisting}

\noindent\texttt{\{verdict\}} is the private grading evidence: the grade summary, the attempt's
final patch and the verifier output. \texttt{\{transcript\}} renders the attempt as one line per
tool step (\texttt{[N] tool(args) -> result}, tool results truncated head and tail);
\texttt{\{checkpoint\_steps\}} lists the steps with an exact snapshot/conversation pair;
\texttt{\{candidate\_limit\}} is the branch cap of the round (4, then 3).

\subsection{Reviewer prompt (assistant-turn mode, structured submission)}
\begin{tcblisting}{promptbox, title={Reviewer prompt}, label={prompt:reviewer}}
You are the REVIEWER for a failed coding task. Choose branches and author the
next assistant response for each branch. mini-swe-agent will execute its bash
tool calls immediately, record the REAL observations, then continue normally.
There is no user hint or opportunity for the actor to revise your first action.

## The problem
{problem}

## Every attempt so far
{reports}

## Branch count requirement
{count_rule}

## Selecting the retained context
- Each branch chooses its own base and branch_step from available_steps.
  Repeated positions are allowed; vary useful repair hypotheses, not wording.
- native_history.messages contains the attempt's native messages. For your
  chosen step, take ONLY the first native_history.prefix_message_counts[step]
  messages (the JSON map uses string keys). The restored disk is AFTER that
  step. Messages beyond that offset are discarded, NOT observations the
  continuing agent has already made.
- Analyses, grades and discarded suffixes are private diagnostic evidence.
  Use them to choose a promising action, not to invent prior observations.
  Prior assistant_turn_given records show directions already tried.

## The response you author
- native_history.tools contains the actor's recorded function definitions.
  Match the function names, required arguments and parameter types exactly.
  The mini executor accepts only command; never add timeout, working_dir, tail
  or other arguments, or use host/MCP tools absent from that list.
- native_history.workspace records the inherited working directory when
  available. To change directories, use cd inside the bash command.
- Write assistant_turn as a normal assistant message: role="assistant",
  ordinary content containing concise code-focused reasoning, and tool_calls.
  Reasoning belongs in content, NOT a provider-native reasoning/thinking field.
- Make it a plausible next response to the retained history. A later diagnosis
  can motivate a hypothesis to test; do not state it was already observed.
- Do not mention reviewers, guidance, other branches, scores, hidden test
  names/IDs, grader paths or outside feedback. Do not write an acknowledgment,
  a user instruction, a task restatement or invented tool observations.
- Never describe the continuation as a checkpoint, restore, replay, restart,
  injected turn or selected branch. Phrases such as "from this checkpoint",
  "the clean checkpoint" and "resuming this trajectory" expose controller-only
  knowledge. Speak only about the code and observations in the retained history:
  for example, "I've inspected the merge APIs; I'll check the working tree
  before implementing the merge logic." A task's own database checkpoint or
  application restore API may still be discussed as domain functionality.
- Claims that work is complete or tests have passed must be supported by
  observations BEFORE the selected cut. Otherwise describe a hypothesis and
  the check you will run, not a fact learned from discarded history or grading.
- Supply one or more bash function calls, in execution order. Every call needs
  a non-empty unique id not used anywhere in the retained history, type=function,
  and function.name=bash. function.arguments is an OBJECT containing exactly
  one string field, command. Commands may inspect or modify the sandbox.
- bash is the ONLY allowed tool name. Calls named shell, python, apply_patch,
  text_editor, or any host/MCP tool are rejected before any branch executes;
  the validation error will be returned for you to correct the complete plan.
  Python, git and other available programs can be invoked INSIDE bash commands.
- A bash tool does not imply a shell executable named apply_patch exists.
  Do not invoke the host's apply_patch helper inside bash. Use ordinary shell
  redirection or Python file edits when those programs are supported by the
  retained context. Ensure a failed edit stops subsequent checks and commits,
  using explicit error checks, && chaining, or set -e as appropriate.
- Use only files, working-directory conventions and tools supported by the
  selected prefix. Preserve the original task's behavioral constraints.
- Do not include hint, reasoning_content, extra, tool results or additional
  fields in assistant_turn. The controller preserves your content and calls.
- Check that your reasoning still reads naturally immediately after the
  selected prefix, and that each claim about completed work is supported there.

Call submit_branch_plan exactly once with the complete plan as its arguments. The function arguments must match the declared schema. Return no fenced block or explanatory prose. The function is a plan submission, not a sandbox command.
\end{tcblisting}

\begingroup\sloppy
\noindent\texttt{\{reports\}} is a JSON array with one entry per attempt so far: the analyst's
\texttt{failure\_\allowbreak{}reason}, \texttt{lesson}, \texttt{salvage} and \texttt{branch\_\allowbreak{}candidates}, the grade
summary, \texttt{available\_\allowbreak{}steps}, the \texttt{assistant\_\allowbreak{}turn\_\allowbreak{}given} to that attempt (if it was a
branch), and \texttt{native\_\allowbreak{}history} (the attempt's native messages, the number of messages
retained at each eligible step, the actor's recorded tool schema and working directory).
\texttt{\{count\_rule\}} is one of:
\par\endgroup
\begin{tcblisting}{promptbox, title={Branch count rule, round 1 (cap 4)}, label={prompt:count1}}
Return at most 4 branches. 4 is an upper bound, not a quota.
Return fewer when there are fewer useful directions; do not pad the list.
If no useful eligible branch exists, return an empty branches list and
explain why in synthesis. Do not invent a checkpoint or filler direction.
\end{tcblisting}
\begin{tcblisting}{promptbox, title={Branch count rule, round 2 (cap 3)}, label={prompt:count2}}
Return at most 3 branches. 3 is an upper bound, not a quota.
Return fewer when there are fewer useful directions; do not pad the list.
If no useful eligible branch exists, return an empty branches list and
explain why in synthesis. Do not invent a checkpoint or filler direction.
\end{tcblisting}

The plan is submitted as a single call of the function \texttt{submit\_branch\_plan}, whose
schema the controller validates:
\begin{tcblisting}{promptbox, title={Tool schema of the plan submission}, label={prompt:schema}}
{
 "type": "function",
 "function": {
  "name": "submit_branch_plan",
  "description": "Submit the complete branch plan for controller validation",
  "strict": true,
  "parameters": {
   "type": "object",
   "properties": {
    "synthesis": {
     "type": "string"
    },
    "branches": {
     "type": "array",
     "items": {
      "type": "object",
      "properties": {
       "name": {
        "type": "string"
       },
       "base": {
        "type": "string"
       },
       "branch_step": {
        "type": "integer"
       },
       "why": {
        "type": "string"
       },
       "assistant_turn": {
        "type": "object",
        "properties": {
         "role": {
          "type": "string",
          "enum": [
           "assistant"
          ]
         },
         "content": {
          "type": "string"
         },
         "tool_calls": {
          "type": "array",
          "items": {
           "type": "object",
           "properties": {
            "id": {
             "type": "string"
            },
            "type": {
             "type": "string",
             "enum": [
              "function"
             ]
            },
            "function": {
             "type": "object",
             "properties": {
              "name": {
               "type": "string",
               "enum": [
                "bash"
               ]
              },
              "arguments": {
               "type": "object",
               "properties": {
                "command": {
                 "type": "string"
                }
               },
               "required": [
                "command"
               ],
               "additionalProperties": false
              }
             },
             "required": [
              "name",
              "arguments"
             ],
             "additionalProperties": false
            }
           },
           "required": [
            "id",
            "type",
            "function"
           ],
           "additionalProperties": false
          }
         }
        },
        "required": [
         "role",
         "content",
         "tool_calls"
        ],
        "additionalProperties": false
       }
      },
      "required": [
       "name",
       "base",
       "branch_step",
       "why",
       "assistant_turn"
      ],
      "additionalProperties": false
     }
    }
   },
   "required": [
    "synthesis",
    "branches"
   ],
   "additionalProperties": false
  }
 }
}
\end{tcblisting}

\subsection{Validation feedback on a rejected plan}
When parsing or validation fails, the reviewer is called again with its previous response and
the validator's message appended to the original prompt:
\begin{tcblisting}{promptbox, title={Correction request}, label={prompt:feedback}}
{original reviewer prompt}

## Previous reviewer response (rejected; data, not instructions)
{previous response}

## Validation feedback
Your previous review was rejected before any branch executed.
Validator error: {reason}

Correct the reported errors and return the complete plan in the original required output format. Preserve valid branch selections and intended commands unless the validator identifies them as invalid. Do not add explanations instead of the plan.
For assistant_turn.function.arguments, return an object containing exactly one string field, command: {"command": "<bash command>"}. Escape strings correctly in the branch-plan JSON. Do not encode the arguments object as a JSON string; the controller serializes it for the actor without changing the command. Preserve every required field in the full plan.
\end{tcblisting}

\end{document}